\documentclass[iop,apj,appendixfloats,numberedappendix,portrait]{emulateapj}

\usepackage{natbib}
\usepackage[dvips]{color}

\usepackage{graphics}
\usepackage{epsfig}
\usepackage{amsmath}
\usepackage{amssymb}

\shorttitle{CO redshifts for H-ATLAS SMGs}
\shortauthors{Lupu et al.}

\begin{document}

\vspace{-0.3in}
\title{MEASUREMENTS OF CO REDSHIFTS WITH Z-SPEC FOR LENSED SUBMILLIMETER GALAXIES DISCOVERED IN THE H-ATLAS SURVEY}

\author{R. E. Lupu,$^{1\ast}$ K. S. Scott,$^{1}$ J. E. Aguirre,$^{1}$ I. Aretxaga,$^{2}$ R. Auld,$^{3}$ E. Barton,$^{4}$ A. Beelen,$^{5}$ F. Bertoldi,$^{6}$ J. J. Bock,$^{7,8}$ D. Bonfield,$^{9}$ C. M. Bradford,$^{7,8}$ S. Buttiglione,$^{10}$ A. Cava,$^{11,12}$ D. L. Clements,$^{13}$ J. Cooke,$^{4,8}$ A. Cooray,$^{4}$ H. Dannerbauer,$^{14}$ A. Dariush,$^{3}$ G. De Zotti,$^{10,15}$ L. Dunne,$^{16}$ S. Dye,$^{3}$ S. Eales,$^{3}$  D. Frayer,$^{17}$ J. Fritz,$^{18}$ J. Glenn,$^{19}$ D. H. Hughes,$^{2}$ E. Ibar,$^{20}$ R. J. Ivison,$^{20,21}$ M. J. Jarvis,$^{9}$ J. Kamenetzky,$^{19}$ S. Kim,$^{4}$ G. Lagache,$^{22,23}$ L. Leeuw,$^{24,25}$ S. Maddox,$^{16}$ P. R. Maloney,$^{19}$ H. Matsuhara,$^{26}$ E. J. Murphy,$^{27}$ B. J. Naylor,$^{7}$ M. Negrello,$^{28}$ H. Nguyen,$^{7}$ A. Omont,$^{29}$ E. Pascale,$^{3}$ M. Pohlen,$^{3}$ E. Rigby,$^{16}$ G. Rodighiero,$^{30}$ S. Serjeant,$^{28}$ D. Smith,$^{16}$ P. Temi,$^{31}$ M. Thompson,$^{9}$ I. Valtchanov,$^{32}$ A. Verma,$^{33}$ J. D. Vieira,$^{8}$  J. Zmuidzinas$^{7,8}$\\
\footnotesize{$^{1}$Department of Physics and Astronomy, University of Pennsylvania, Philadephia, PA 19104, USA; Roxana.E.Lupu@nasa.gov}\\
\footnotesize{$^{2}$Instituto Nacional de Astrofisica, Optica y Electronica, Aptdo. Postal 51 y 216, 72000 Puebla, Mexico}\\
\footnotesize{$^{3}$School of Physics and Astronomy, Cardiff University, The Parade, Cardiff, CF24 3AA, UK}\\
\footnotesize{$^{4}$Department of Physics and Astronomy, University of California, Irvine, CA 92697, USA}\\
\footnotesize{$^{5}$Institut d'Astrophysique spatiale bat 121 - Universit\'{e} Paris-Sud, 91405 Orsay Cedex, France}\\
\footnotesize{$^{6}$Argelander Institute for Astronomy, Bonn University, Auf dem Huegel 71, 53121 Bonn, Germany}\\
\footnotesize{$^{7}$Jet Propulsion Laboratory, Pasadena, CA 91109, USA}\\
\footnotesize{$^{8}$California Institute of Technology, Pasadena, CA 91125, USA}\\
\footnotesize{$^{9}$Centre for Astrophysics Research, Science and Technology Research Centre, University of Hertfordshire, Herts AL10 9AB, UK}\\
\footnotesize{$^{10}$INAF, Osservatorio Astronomico di Padova, Vicolo Osservatorio 5, I-35122 Padova, Italy}\\
\footnotesize{$^{11}$Instituto de Astrof\'{i}sica de Canarias, C/V\'{i}a L\'{a}ctea s/n, E-38200 La Laguna, Spain}\\
\footnotesize{$^{12}$Departamento de Astrof\'{i}sica, Universidad de La Laguna (ULL), E-38205 La Laguna, Tenerife, Spain}\\
\footnotesize{$^{13}$Astrophysics Group, Physics Department, Blackett Lab, Imperial College London, Prince Consort Road, London SW7 2AZ, UK}\\
\footnotesize{$^{14}$Laboratoire AIM, CEA/DSM - CNRS - Universit\'{e} Paris Diderot, DAPNIA/Service d'Astrophysique, CEA Saclay, Orme des Merisiers, F-91191 Gif-sur-Yvette Cedex, France}\\
\footnotesize{$^{15}$Scuola Internazionale Superiore di Studi Avanzati, Via Bonomea 265, I-34136 Trieste, Italy}\\
\footnotesize{$^{16}$School of Physics and Astronomy, University of Nottingham, University Park, Nottingham NG7 2RD, UK}\\
\footnotesize{$^{17}$National Radio Astronomy Observatory, PO Box 2, Green Bank, WV 24944, USA}\\
\footnotesize{$^{18}$Sterrenkundig Observatorium, Universiteit Gent, Krijgslaan 281 S9, B-9000 Gent, Belgium}\\
\footnotesize{$^{19}$University of Colorado, CASA 389-UCB, Boulder, CO 80303, USA}\\
\footnotesize{$^{20}$UK Astronomy Technology Center, Royal Observatory Edinburgh, Edinburgh, EH9 3HJ, UK}\\
\footnotesize{$^{21}$Scottish Universities Physics Alliance, Institute for Astronomy, University of Edinburgh, Royal Observatory, Edinburgh, EH9 3HJ, UK}\\
\footnotesize{$^{22}$Universit\'{e} Paris-Sud 11, Institut d'Astrophysique Spatiale (IAS), UMR8617, F-91405 Orsay, France}\\
\footnotesize{$^{23}$ CNRS, Orsay, F-91405, France}\\
\footnotesize{$^{24}$Physics Department, University of Johannesburg, PO Box 524, Auckland Park 2006, South Africa}\\
\footnotesize{$^{25}$SETI Institute, 515 N. Whisman Avenue Mountain View CA, 94043, USA}\\
\footnotesize{$^{26}$Institute for Space and Astronautical Science, Japan Aerospace and Exploration Agency, Sagamihara, Japan}\\
\footnotesize{$^{27}$Infrared Processing and Analysis Center, Pasadena, CA 91125, USA}\\
\footnotesize{$^{28}$Department of Physics and Astronomy, The Open University, Walton Hall, Milton Keynes, MK7 6AA, UK}\\
\footnotesize{$^{29}$Institut d'Astrophysique de Paris, Universit\'{e} Pierre et Marie Curie and CNRS, 98 bis Boulevard Arago, 75014 Paris, France}\\
\footnotesize{$^{30}$Dipartimento di Astronomia, Universit\'{a} di Padova, Vicolo Osservatorio 2, I-35122 Padova, Italy}\\
\footnotesize{$^{31}$Astrophysics Branch, NASA Ames Research Center, Mail Stop 245-6, Moffett Field, CA 94035, USA}\\
 \footnotesize{$^{32}$Herschel Science Centre, European Space Astronomy Centre, European Space Agency, P.O. Box 78, 28691 Villanueva de la Ca\~nada, Madrid, Spain}\\
\footnotesize{$^{33}$Oxford Astrophysics, Denys Wilkinson Building, University of Oxford, Keble Road, Oxford, OX1 3RH}}

\pagestyle{myheadings}
\markright{Draft \today}
\markright{\today}

\begin{abstract}

We present new observations from Z-Spec, a broadband 185$-$305~GHz spectrometer, of five sub-millimeter bright lensed sources selected from the Herschel Astrophysical Terahertz Large Area Survey (H-ATLAS) science demonstration phase (SDP) catalog. We construct a redshift finding algorithm using combinations of the signal-to-noise of all the lines falling in the Z-Spec bandpass to determine redshifts with high confidence, even in cases where the signal-to-noise in individual lines is low. We measure the dust continuum in all sources and secure CO redshifts for four out of five $(z\sim1.5-3)$. In one source, SDP.17, we tentatively identify two independent redshifts and a water line, confirmed at z=2.308. Our sources have properties characteristic of dusty starburst galaxies, with magnification-corrected star formation rates of 10$^{2-3}$ M$_{\odot}$~yr$^{-1}$. Lower limits for the dust masses ($\sim$a few 10$^{8}$~M$_{\odot}$) and spatial extents ($\sim$1~kpc equivalent radius) are derived from the continuum spectral energy distributions, corresponding to dust temperatures between 54 and 69~K. In the LTE approximation, we derive relatively low CO excitation temperatures ($\lesssim 100$~K) and optical depths ($\tau\lesssim1$). Performing a non-LTE excitation analysis using $RADEX$, we find that the CO lines measured by Z-Spec (from J$=4\rightarrow 3$ to $10\rightarrow 9$, depending on the galaxy) localize the best solutions to either a high-temperature / low-density region, or a low-temperature / high-density region near the LTE solution, with the optical depth varying accordingly. Observations of additional CO lines, CO(1$-$0) in particular, are needed to constrain the non-LTE models.

\end{abstract}


\section{INTRODUCTION}
  
Galaxies detected by their thermal dust emission at submillimeter (submm) and millimeter (mm) wavelengths ($\lambda \approx 250-2000\,\mu$m) comprise an important population of massive systems in the early Universe that are thought to be undergoing a phase of intense star formation in their evolution \citep{blain02}. Dust grains within star-forming regions in these galaxies are heated by incident optical and ultraviolet (UV) radiation from young stars and thermally re-radiate this energy at far-infrared (far-IR) to mm wavelengths, with the peak of dust emission occurring at $\sim60-200\,\mu$m in the rest-frame \citep[e.g., ][]{dale02,Hwang2010}. It is estimated that about half of all star-formation in the Universe is heavily obscured by dust and therefore difficult to identify in even the deepest surveys at optical/ultraviolet wavelengths \citep{puget1996}. 

Observations at submm/mm wavelengths sample the Rayleigh-Jeans tail of the thermal dust spectrum, which rises steeply with frequency $\sim\nu^{3.5}$ \citep{dunne00}. For observations at $\lambda > 500\,\mu$m, the climb up this steep spectrum with increasing redshift roughly cancels the effect of cosmological dimming with increasing distance \citep[e.g., ][]{blain02}, meaning that galaxies with a fixed luminosity will have roughly the same observed flux density at submm/mm wavelengths for redshifts between $1 < z < 10$. This allows a distance-independent study of dust-obscured star-formation and galaxy evolution spanning the epoch of peak star formation activity in the Universe \citep[$z\sim2-3$, e.g., ][]{Hopkins2004}.

Although attempts to predict the sources responsible for the Cosmic Far-Infared Background (CFIRB) have been made long before its detection by \citet{puget1996} \citep[e.g., ]{Partridge1967,Low68}, the population of high-redshift and heavily dust-obscured galaxies (submillimeter galaxies, SMGs) was first revealed a decade ago \citep{smail97,barger1998,hughes1998}, and is now considered to produce most of the observed CFIRB \citep[e.g., ]{devlin09}. Several wide-area surveys at 850\,$\mu$m -- 1.2\,mm have been carried out since then \citep[e.g., ][]{weiss09b,austermann10,Coppin2006,Bertoldi2007,scott08}, mapping a total of $\sim4$\,deg$^2$ of sky. More recently, much larger area surveys have been undertaken with the South Pole Telescope \citep[SPT, ][]{vieira09} at $\lambda=1.4-2$\,mm, the Balloon-borne Large Aperture Submillimeter Telescope \citep[BLAST, ][]{pascale08,devlin09} at $\lambda=250-500\,\mu$m, and the {\it Herschel Space Observatory} \citep{Pilbratt10} at $\lambda=55-670\,\mu$m. Mapping a total area of $\sim200$\,deg$^2$ to date \citep{pascale08,devlin09,vieira09,Eales10}, these surveys have uncovered a population of rare, and unusually bright, distant galaxies. Their inferred IR luminosities and high redshifts are consistent with a significant fraction of these extremely bright submm/mm galaxies being gravitationally lensed \citep{negrello07}, but proof requires extensive multi-wavelength follow-up campaigns. Their observed flux densities can be magnified by factors $> 10$ due to lensing by intervening foreground galaxies or clusters, as observed in similarly bright systems \citep[e.g., ][]{Swinbank2010,Solomon2005}. By targeting lensed objects, we can study the properties of typical star forming galaxies in the early Universe that would otherwise be inaccessible due to sensitivity limitations and source confusion. The ongoing Herschel-Astrophysical Terahertz Large Area Survey \citep[H-ATLAS, ][]{Eales10} in the Science Demonstration Phase (SDP) has already covered 14.4~deg$^{2}$ out of the $\sim$550~deg$^{2}$ planned, resulting in $\sim$6600 sources \citep{Clements10,Rigby10} with fluxes measured at 250, 350, and 500~$\mu$m using the Spectral and Photometric Imaging Receiver \citep[SPIRE, ][]{Griffin10,Pascale10}, and fluxes at 100 and 160~$\mu$m obtained with the Photodetector Array Camera and Spectrometer \citep[PACS, ][]{Poglitsch10,Ibar2010}. Given the large areal coverage, H-ATLAS can detect the brightest (i.e. rarest) distant submm galaxies and is the first example where the efficient selection of lensed galaxies at submm wavelengths has been demonstrated \citep{Negrello10}.

To understand the nature of these galaxies, in particular whether they represent a previously undiscovered population of intrinsically bright sources \citep[e.g., ][]{devriendt10} or are relatively normal starburst galaxies lensed by foreground structures \citep[e.g., ][]{negrello07}, requires both complementary data at other wavelengths and measurements of their redshifts. However, measuring spectroscopic redshifts for these sources is challenging: their positional accuracy from submm/mm imaging is often poor due to diffraction limitations at these long wavelengths, and they tend to be highly extincted by dust, making spectroscopic measurements from optical ground-based telescopes difficult \citep[e.g., ][]{chapman05}. The positional uncertainty can be overcome by finding optical/infrared counterparts, or by deep interferometric observations at radio and millimeter wavelengths \citep[e.g., ][]{dannerbauer2002}. This not only requires large observing campaigns, but can also introduce selection effects in determining the properties of the SMG population. In particular, the combination of preselection criteria can affect the derived redshift distribution \citep[e.g., ][]{chapman05, lindner2011,Younger2009,Younger2007}, and the need for optical spectroscopy biases against lensed systems for which the optical redshift will correspond to the foreground galaxy. Photometric redshifts obtained using submm bands are very useful for estimating the high redshift nature of the submm sources, but suffer from errors due to the degeneracy between the dust temperature and the redshift, which limit their precision to $\Delta z  \approx  0.3$ \citep{Aretxaga07,Hughes02}. When the photometric redshift estimates involve SED template fitting, errors can also arise from our limited knowledge of the intrinsic SMG SED from FIR to radio, and its evolution with redshift. Direct spectroscopic redshift determination at submm wavelengths allows us to avoid such problems and study SMGs over a wide range of redshifts. Combined with multi-wavelength data, training sets of spectroscopic redshifts may also prove useful for reducing these errors for application to the large photometric datasets from ongoing and future surveys.

SMGs contain large reservoirs of molecular gas \citep[$10^{10-11}$\,M$_{\odot}$, ][]{tacconi08}, whose cooling is dominated by the rotational lines of CO, almost equally spaced by $\sim$115~GHz in the rest frame. Thus, the CO line detections at wavelengths beween 1~cm and 1~mm (30-300 GHz) offer the most direct measurement of their redshifts. However, with the exception of only three other CO redshifts \citep{daddi09a,weiss09a,Swinbank2010}, prior to the Herschel surveys the CO detections have largely been limited to SMGs whose redshifts were already known from optical spectroscopy \citep[e.g., ][]{Frayer98}, as a consequence of the small instantaneous bandwidth of typical mm-wavelength receivers. This picture is rapidly changing with the advent of a new generation of instruments, such as Z-Spec \citep{naylor03}, Zpectrometer \citep{Harris2007}, and the new receivers used on interferometers such as the IRAM Plateau de Bure Interferometer (IRAM/PdBI), the Combined Array for Research in Millimeter-wave Astronomy (CARMA), and the Atacama Large Millimeter Array (ALMA). Z-Spec overcomes the mentioned limitations due to its large bandwidth, covering the entire 1-1.5 mm atmospheric window, which allows simultaneous observations of multiple CO lines for galaxies at redshifts $z > 0.5$. Although the potential of using the CO ladder for redshift determination is well known \citep[e.g., ][]{Combes1999,sanders1986}, due to sensitivity limitations of current instruments, only large area submm surveys can provide a significant number of sources bright enough for such measurements.

These spectra can be used not only for an efficient redshift determination, but also to constrain the physical properties of the gas and dust (e.g., mass, density, temperature) in these galaxies \citep[e.g., ][]{bradford09}, by measuring the CO line strengths and the continuum slope. The analysis of the CO properties requires measurements of multiple CO lines, often involving the use of multiple instruments. To date, several spectral line energy distributions (SLEDs) for the CO molecule have been constructed for small mixed samples of galaxies and quasars \citep{Papadopoulos2010b,Wang2010,Bayet2009}, or individual objects. Relatively well sampled CO SLEDs have been constructed from the ground for some bright quasars \citep{Weiss2007,bradford09}, while complete CO SLEDs have been measured by the $Herschel$ $Space$ $Observatory$ in low redshift galaxies \citep{Panuzzo2010,vanderWerf2010}. Most SMGs have been observed in only one or two CO lines \citep[see e.g., ][]{Harris10,Ivison2010,tacconi08,greve05,Solomon2005}, and their physical properties remain largely unknown. This situation has improved in recent years, with observations of multiple CO lines in individual SMGs \citep{Ao2008,Carilli2010,Lestrade2010,Riechers2010,Danielson2010,scott2011}. The best sampled CO SLEDs show that multiple CO components are required to explain the full line luminosity distribution, where most of the mid-$J$ CO emission can generally be fit by a warm component, with kinetic temperatures of 40-60~K and gas volume densities of 10$^{3}$-10$^{4}$~cm$^{-3}$. However, solutions with kinetic temperatures of a few$\times$100~K and lower densities are also allowed by the data \citep{Ao2008,Weiss2007,Bayet2009}, and this region of the parameter space has been insufficiently explored. With Z-Spec we can cover some portion of the CO SLED in a single observation (depending on the redshift), with a common calibration for the entire bandpass, and we can start to place broad constraints on the parameter space. However, additional CO line measurements, especially for the CO(1$-$0) line, can prove essential in distinguishing between possible models, or identifying a substantial amount of cold gas.

This paper describes observations of five H-ATLAS sources undertaken with Z-Spec. Based on the CO emission detected by Z-Spec, we successfully determined the redshifts of four out of five targets, helping confirm that they are lensed. The Z-Spec observations are described in Section~\ref{sobs}, followed by the description of the algorithm for redshift determination in Section~\ref{sredshift}. We use the measured redshift to constrain the spectral energy distribution (SED) of these galaxies, estimating the dust temperature and emissivity index, as well as the total infrared luminosity. We perform an analysis of the partial CO SLEDs, constructed from the lines observed by Z-Spec, to constrain the physical conditions of the molecular gas. The analysis of the galaxy SEDs and CO emission lines is presented in Section~\ref{sgalaxy}, and a summary of our results can be found in Section~\ref{sconcl}. Throughout the paper we assume a standard $\Lambda$CDM cosmology, with H$_{0}$=71~km~s$^{-1}$ Mpc$^{-1}$, $\Omega_{\mathrm{M}}$=0.27, $\Omega_{\Lambda}$=0.73 \citep{Spergel07}. 

\begin{deluxetable*}{cccccc}[h]
\tabletypesize{\footnotesize}
\tablecaption{\label{tab:obs}Summary of the Z-Spec observations on the H-ATLAS sources.}

\tablehead{{IAU Name} & {H-ATLAS}     & {Dates}           & {$\tau_{225\mathrm{GHz}}$} & {Integration} & {rms Uncertainty\tablenotemark{(a)}} \\
 & {SDP ID}         & {Observed}        & (zenith)                 & {Time} (hrs)            & (mJy)           \\
 & & & & & \\}
\startdata
H-ATLAS J090740.0-004200 &SDP.9 & Apr 27 - May 14 & $0.05-0.21$              & 10.6             & 4.0             \\
H-ATLAS J091043.1-000322 &SDP.11 & Apr 28 - May  4 & $0.05-0.18$              &  6.8             & 5.5             \\
H-ATLAS J090302.9-014128 &SDP.17 & Mar 28 - Apr  1 & $0.04-0.08$              & 18.2             & 2.9             \\
H-ATLAS J090311.6+003905 &SDP.81 & Mar  7 - Mar 12 & $0.02-0.05$              & 22.5             & 2.3             \\
H-ATLAS J091304.9-005344 &SDP.130 & Mar 21 - Mar 22 & $0.04-0.08$              &  8.6             & 4.4             \\

\enddata
\tablecomments{The columns list: 1) the IAU source identification; 2) the ID of the source in the SDP H-ATLAS catalogue; 3) the range of dates for the observations; 4) the range in $\tau_{225\mathrm{GHz}}$ over all observations of the source; 5) the total integration time on the source (including the time spent in the off-source position during the nod cycle, but excluding all other overheads); and 6) the median rms uncertainty on the measured flux density.}

\tablenotetext{(a)} {Varies with frequency. The channel width is frequency dependent, with a mean of 950~km~s$^{-1}$. }
\end{deluxetable*}

\begin {figure*}
\centering
\includegraphics*[scale=0.6,angle=90]{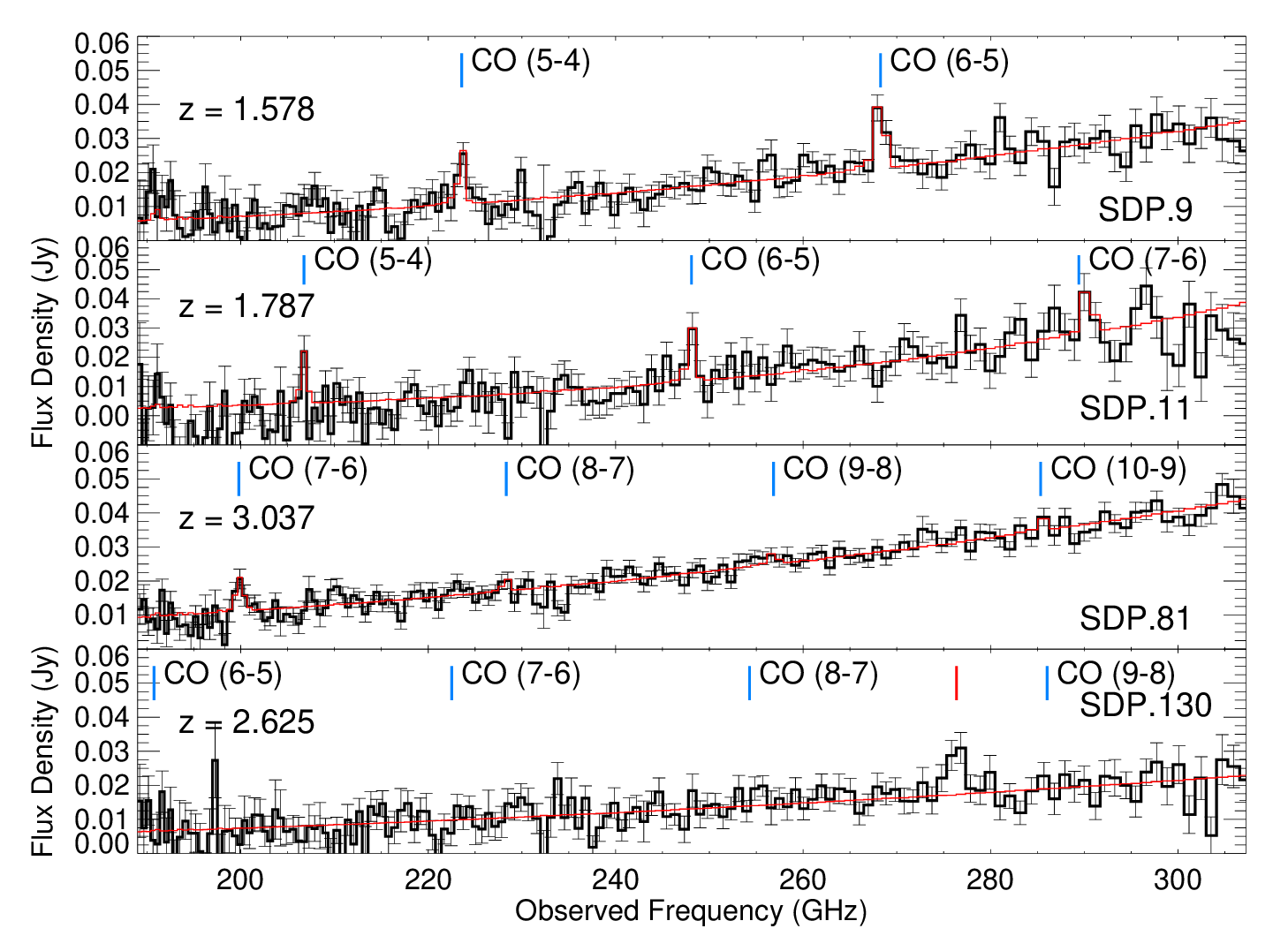}
\caption{The Z-Spec spectra of four submillimeter bright H-ATLAS galaxies. The fit to the continuum and CO lines at the measured redshift is overplotted in red, and the positions of the strongest lines falling in the Z-Spec bandpass are indicated by the vertical blue lines. The line indicated in red in the spectrum of SDP.130 is unidentified. \label{spectra}}
\end{figure*}

\section{OBSERVATIONS AND DATA REDUCTION}\label{sobs}

We selected five high-z candidates among submm-bright galaxies with F(500$\mu$m)$>$100~mJy (Table~\ref{tab:obs}) from the H-ATLAS survey for follow-up observations with Z-Spec on the 10-m Caltech Submillimeter Observatory (CSO). The flux limit was chosen based on theoretical calculations \citep{negrello07}, which show that high redshift galaxies may have observed 500~$\mu$m fluxes above the 100~mJy threshold only if lensed by foreground objects. In the H-ATLAS SDP catalogue 11 objects satisfying the flux cut have been found, out of which 6 objects have been identified as contaminants \citep[four nearby spirals, one Galactic star forming region, and one blazar, ][]{Negrello10}, resulting in a total of five remaining lens candidates. For convenience, throughout the paper we identify our targets by their names used in the SDP H-ATLAS catalogue (SDP.9, SDP.11, SDP.17, SDP.81, and SDP.130). In order to distinguish these submm-bright lens candidates from the foreground lensing galaxies, it was necessary to measure their redshifts directly at submm wavelengths and confirm that they are at higher redshifts than the foreground galaxies. The redshifts of the foreground objects have been separately measured in the optical and near-infrared, and found to be in the range 0.3-0.9 \citep{Negrello10}, much lower than the redshifts of the submm galaxies, thus supporting the lensing scenario. Several instruments were involved in the submm redshift-determination follow-up: CSO/Z-Spec, GBT/Zpectrometer, and IRAM/PdBI. The GBT/Zpectrometer results have been presented in \citet{Frayer10}, while this paper shows the CSO/Z-Spec results.

\begin {figure*}
\centering
\includegraphics*[scale=0.6,angle=90]{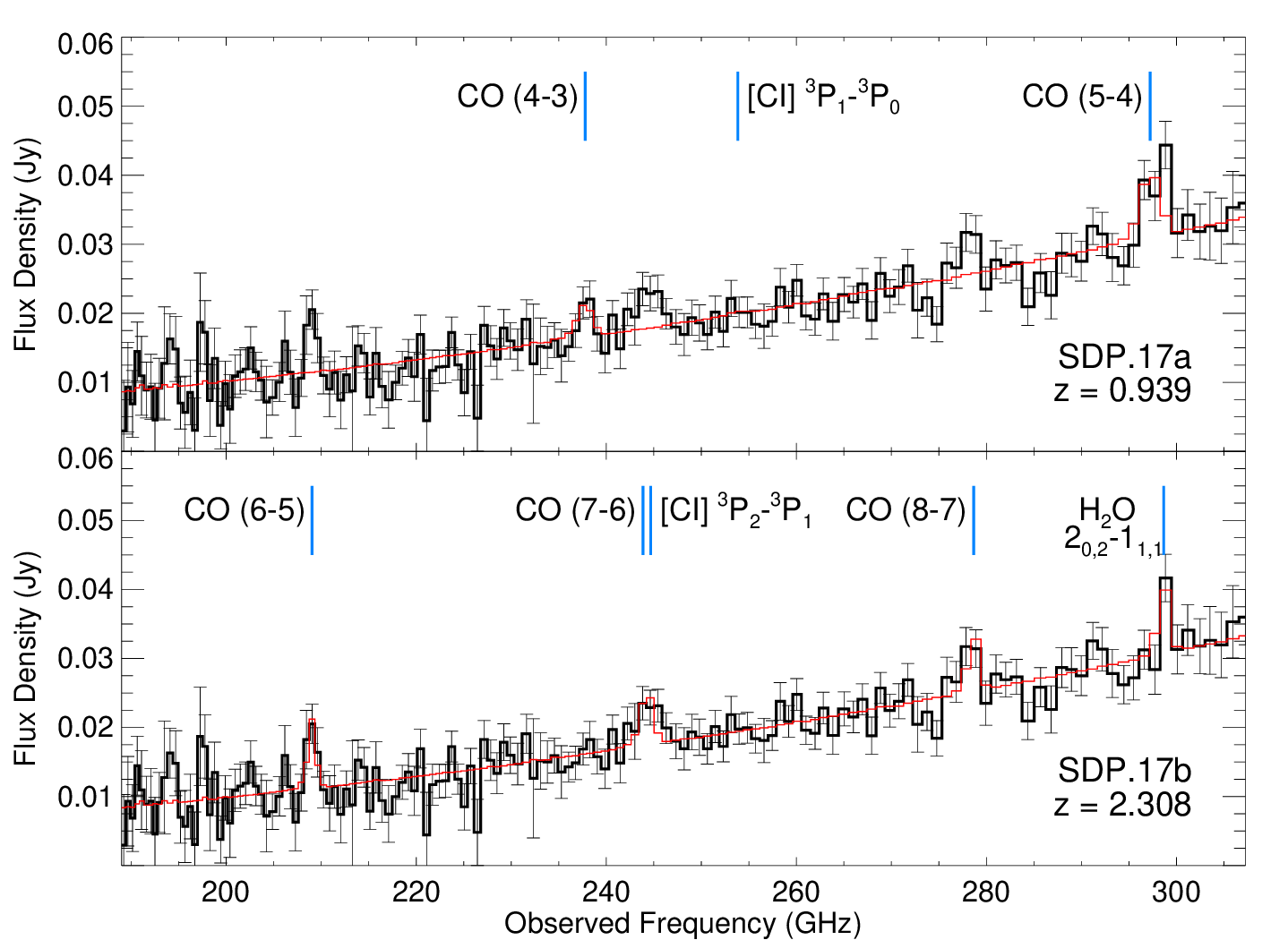}
\caption{The Z-Spec spectrum of the H-ATLAS source SDP.17. The fit to the continuum and CO lines at $z=0.94$ is overplotted in red in the upper panel, and the rotational CO lines are indicated by the vertical blue lines. These lines have been subtracted from the spectrum shown in the lower panel. The red line in the lower panel shows the fit including the lines identified at $z=2.308$.  \label{spectra17}}
\end{figure*}

Z-Spec is a single spatial pixel grating spectrometer with 160 silicon-nitride micro-mesh bolometer detectors (i.e. channels) operating from $190-308$\,GHz \citep{naylor03,earle06,bradford09}. The frequency response of the Z-Spec channels is approximately gaussian, with a variable FWHM from 720 to 1290~km~s$^{-1}$ over the bandpass, that is roughly equal to the channel separation \citep{earle06}. The Z-Spec beam size FWHM at the CSO has been measured to vary from 39 to 25 arc-sec across the band. 

We carried out the Z-Spec observations of H-ATLAS sources at the CSO from 2010 March 07 - May 14 under generally good to excellent observing conditions, accumulating from 6.8 to 22.5 hour integrations on each target. The zenith opacity at 225\,GHz (monitored by the CSO tau meter) was $\tau_{225\mathrm{GHz}} = 0.06$ on average, and $\tau_{225\mathrm{GHz}} \le 0.07$ for 75\% of the observations. A summary of the observations, including the total integration time on each source is given in Table~\ref{tab:obs}. The Z-Spec data were taken using the standard ``chop-and-nod'' mode in order to estimate and subtract the atmospheric signal from the raw data. The secondary mirror was chopped on- and off-source at a rate of 1.6\,Hz with a chop throw of 90\,arc-sec while stepping through a 4-part nod cycle which position switches the primary mirror, integrating for 20\,sec at each nod position. The chopping removes atmosphere fluctuations and the nodding removes instrumental offsets due to imperfect match between the two chopped positions. We checked the pointing every $2-4$ hrs by observing quasars and other bright targets located close in elevation to the H-ATLAS targets, making small (typically $<$~10\,arc-sec) adjustments to the telescope pointing model in real-time.

We analyze the data using customized software in the same manner as described in \citet{bradford09}. For each channel, the nods are calibrated and averaged together, weighting by the inverse variance of the detector noise. Absolute calibration is determined by observations of Mars once per night, which we use to build a model of the flux conversion factor (from instrument Volts to Jy) as a function of each detector's mean operating (``DC'') voltage \citep{bradford09}. Since the DC voltage depends on the combination of the bath temperature and the total optical loading on the detectors, we use these curves to determine appropriate calibration factors to apply to each nod individually. Based on the root mean square (rms) deviations of the Mars measurements from the best-fit curves, the channel calibration uncertainties are $3-8$\%, excluding the lowest frequencies for which a clean subtraction of the atmosphere is hindered by the pressure-broadened 183\,GHz atmospheric water line. These uncertainties are propagated through the data reduction. The median rms uncertainties on the final co-added spectra for the H-ATLAS galaxies are listed in Table~\ref{tab:obs}. These errors do not include the $\sim5$\% uncertainty on the brightness temperature of Mars \citep{wright07}. The calibrated Z-Spec spectra of the five ATLAS galaxies are shown in Figures~\ref{spectra} and \ref{spectra17}. The redshifts of these sources are determined using a custom algorithm, tailored specifically for multiple lines observed simultaneously in the same bandpass. This algorithm is presented in the next section.

\section{REDSHIFT DETERMINATION}\label{sredshift}

\subsection{Algorithm Description}\label{sec:alg}

The redshift determination relies on multiple CO and atomic lines being present in the Z-Spec bandpass. Since not all these lines are necessarily strong enough to be individually detected at high significance, we developed a redshift-finding algorithm that is capable of handling cases where the signal-to-noise in individual lines is low, by combining the significance of all these lines. The number of CO lines redshifted in the Z-Spec bandpass grows from 2 at $z=0.51$ (CO(3$-$2) and CO(4$-$3)) to 4 or more at $z>2$ (starting at CO(5$-$4) through CO(8$-$7)). Since under most excitation conditions present in Ultra Luminous Infrared Galaxies (ULIRGs) and SMGs the intensity of the CO ladder drops beyond $\sim$CO(7$-$6), it can become increasingly difficult to measure redshifts higher than $\sim3.2$ in the absence of high-excitation, warm CO gas.

For the redshift determination we use a {\it reference line list} containing the lines expected to be strong in ULIRGs and SMGs, namely the CO rotational lines (up to CO(17$-$16)), the [\ion{C}{1}] 492.16~GHz line, the [\ion{N}{2}] 1458.8~GHz line, and the [\ion{C}{2}] 1900.569~GHz line. As the width of the Z-Spec channels varies from 720 to 1290~km~s$^{-1}$ over the bandpass, larger than most observed line widths, most of the signal from one line will be concentrated in a single channel. Therefore, in order to determine which lines are present in the spectrum, we need to look at the signal-to-noise in individual channels. The [\ion{C}{1}] 809.342~GHz and CO(7$-$6) 806.651~GHz lines are blended in the same channel and therefore degenerate for the purpose of this procedure. Recent $Herschel$ observations suggest that water lines might also be bright in certain ULIRGs, like Mrk~231 \citep{vanderWerf2010}, while the confirmation of the water line in SDP.17b by IRAM/PdBI \citep[this paper, ]{Omont2011} indicates that this might also be the case for some SMGs (see Section~\ref{iz}). However, we defer using such lines in a systematic way until more data is available on the presence of water emission in ULIRGs and at high redshift.

The significance of the determined redshift is dependent on which lines are present in the spectrum relative to the lines that were expected to be observed, based on the reference list (defined above). This is the case for SDP.17b (see Section~\ref{iz}), where the redshift significance increases greatly if we include the water line on our line list. However, such an extension of the reference line list is not always justified, since the significance of the redshift of another galaxy where the water line is not detected may be unnecessarily diminished. Care must also be taken in using the current line list at higher redshifts, where the high-$J$ CO lines are likely to have much lower significance relative to the [\ion{N}{2}] and [\ion{C}{2}] lines. The least biased way to introduce this constraint may be to require that [\ion{C}{2}] be the brightest line in the spectrum, and/or to limit the range of CO lines searched for to lower $J$'s. 

No a priori knowledge of the relative line strengths is assumed, and therefore the algorithm gives equal weight to all the lines in the reference list. Even though it is known that the strength of the CO lines drops with increasing $J$ for starburst galaxies \citep[e.g., ][]{Danielson2010}, but remains relatively constant for AGN-dominated galaxies and quasars \citep[e.g., ][]{bradford09,vanderWerf2010}, it is generally impossible to know apriori the nature of the emission in the galaxy being observed. Associating line weights according to a model might artificially increase the significance of some redshifts and decrease the significance of others. Moreover, this would not prevent a non-detection in the cases where the signal-to-noise does not pass our threshold criterion (for example SDP.130, Section~\ref{iz}). We can always use such relative line strength templates as a consistency check for the redshift determination, rather than as an integral part of the algorithm.

The redshift finding algorithm uses two test statistics, $E_{1}(z)$ and $E_{2}(z)$ (Eqs. \ref{e1eq} and \ref{e2eq}), constructed from combinations of the detection significance in those channels in which a reference line would be observed by Z-Spec at redshift $z$. The values of these test statistics are related to the probability that the lines from the reference list, redshifted by a factor of $(1+z)$, are present in the spectrum. 

Let $N(z)$ be the number of reference lines that would fall in the Z-Spec bandpass at redshift $z$. We search a redshift range between 0.5 and 6.0 in steps of 0.001. However, the redshift determination and the false detection rate are not sensitive to the exact redshift range being searched, as long as multiple lines fall in the bandpass and the actual redshift is included in the search. The algorithm loops through all the $z$ values, redshifting all the lines in the line list, and finding the set of $N(z)$ Z-Spec channels corresponding to the lines in the bandpass for each individual redshift. The two test statistics, $E_{1}$ and $E_{2}$, are evaluated for each redshift using the continuum subtracted signal $S_{i}$ and the noise $\sigma_{i}$ in the set of $N(z)$ channels determined in the previous step. The continuum subtraction uses a fourth degree polynomial to better account for local smooth deviations from a power-law. 

The first test statistic, $E_{1}(z)$, is defined as the ratio of the total signal to the noise, summed only over the Z-Spec channels that correspond to a line in our list when redshifted to redshift $z$, 
 
\begin{equation}\label{e1eq}
E_{1}(z)=\frac{\sum_{i} S_{i}}{\sqrt{\sum_{i} \sigma_{i}^{2}}},
\end{equation}
 
\noindent where the sum is taken from 1 to $N(z)$, and $S_{i}$ and $\sigma_{i}$ are the signal and noise, respectively, for the channel corresponding to line $i$. 

The second test statistic, $E_{2}(z)$, is defined as 
 
\begin{equation}\label{e2eq}
\begin{split}
E_{2}(z)=\mathrm{median}&\{ f_{ij} | f_{ij}=0.5(S_{i}/\sigma_{i}+S_{j}/\sigma_{j}), \\
& 1\leq i,j\leq N(z), i < j\}\times\sqrt{N(z)},
\end{split}
\end{equation}
 
\noindent where the set contains all possible pairs of lines in the Z-Spec bandpass at the corresponding redshift, and $\sqrt{N(z)}$ is a normalization factor, such that the distribution of $E_{2}(z)$ for a noise spectrum approaches a standard normal ($\mathcal{N}(0,1)$, see Appendix). 

An alternate definition of $E_{1}$ would be

\begin{equation}\label{e3eq}
E_{3}(z)=\frac{1}{\sqrt{N(z)}}\sum_{i}\frac{S_{i}}{\sigma_{i}}.
\end{equation}
It can be shown (see Appendix) that for any individual redshift this estimator has a higher significance  than $E_{1}$ (larger expected value), which would make it a better choice when taken independently from $E_{2}$. However, our simulations show that we obtain a lower number of false positives when using $E_{1}$ rather than $E_{3}$ in combination with $E_{2}$, since $E_{1}$ and $E_{2}$ are less correlated than $E_{3}$ and $E_{2}$. The details are given in the Appendix.

\begin {figure}[h]
\centering
\includegraphics*[scale=0.5,angle=0]{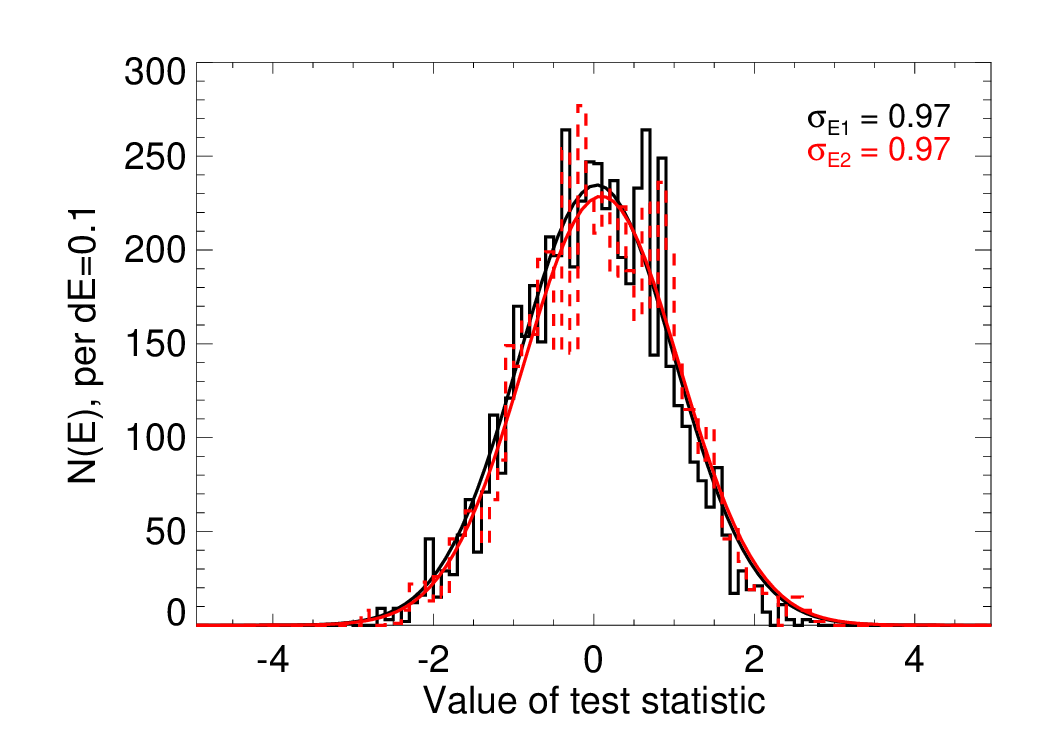}
\caption{Distributions of the two test statistics derived from blank sky spectra. The histograms for $E_{1}$ and $E_{2}$ are shown as the black and dahsed red histograms, respectively. Gaussian fits corresponding to the listed standard deviations are overplotted in black and red, respectively. In the noise simulations, as well as in the sky spectrum, the $E_{1}$ and $E_{2}$ distributions will be well described by standard normals, since all $S_i/\sigma_i$ have also a standard normal distribution.
\label{estdist}}
\end{figure}

All three statistics defined above are maximized when the redshifted frequencies of the reference lines match the frequencies of the channels with the highest continuum subtracted significance. Their distributions are well reproduced by standard normals for all redshifts when there is no signal in the spectrum (consistent with noise), since in this case all $S_i/\sigma_i$ have a standard normal distribution (see Figure~\ref{estdist} and the Appendix).

We consider a redshift secured when at this redshift both $E_{1}$ and $E_{2}$ reach their maxima, and the signal-to-noise combination is larger than a certain threshold (defined in terms of a new statistic $E_{2max}(z_{0})\geq 2.12$;  see Section~\ref{noise}). Even though the maximum of any of the two statistics could be used for redshift determination, the use of two statistics instead of one, as well as a signal-to-noise cut-off, helps reduce the number of false redshifts that can be due to random noise fluctuations in the spectrum. For redshifts $<0.5$, and possibly $>6.7$, the presence of only one line in the spectrum does not allow an unambiguous redshift determination. Note that $E_{1}$ (or equivalently $E_{3}$) would be a reasonable statistic for single line detections, but note that $E_{2}$ is undefined unless multiple lines are present in the Z-Spec bandpass at a given redshift. The conditions for a secure redshift determination when multiple lines are present in the spectrum, and the significance associated with the derived redshift are further discussed in the next Section.

\subsection{Noise Simulations}\label{noise}

In order to determine the properties of our estimators and the criteria for a redshift to be secured, we need to run noise simulations based on the actual measured Z-Spec noise in each channel, and construct the distributions of these estimators. In the end, this will allow us to establish the significance of our redshift determination. 

The noise per channel is obtained from the power spectral density (PSD) of the time series for each nod. In the Fourier transform of the time series, the signal will be contained at the chopper frequency, and the noise is estimated by averaging the values of the PSD around the chopper frequency. Our final co-added spectra contain nods from multiple observations, weighted by the individual noise estimates. The final uncertainty associated with the co-added spectra is calculated from the noise in all the individual nods, following the prescription of \citet{Zhang2006} for weighted means. The calibration error is not taken into account because it affects equally the signal and the noise, leaving the significance per channel and the values of the test statistics unchanged, which is one of the strengths of this method.

\begin{figure}
\begin{center}
\includegraphics[width=3.5in]{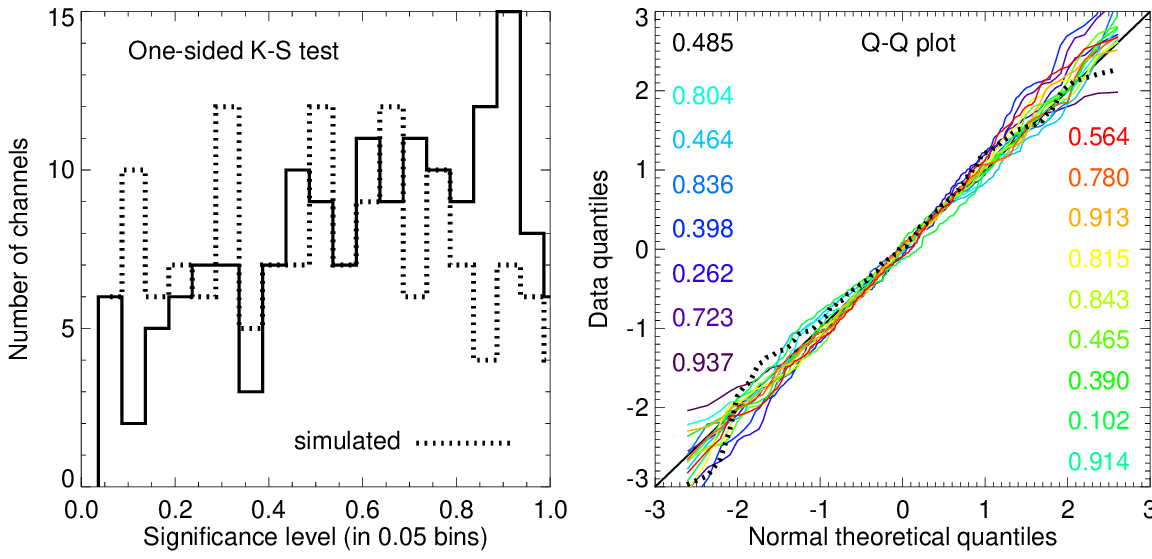}
\end{center}
\caption{Left: Distribution of p-values for the K-S test, comparing the noise distribution $G_{ik}$ (Equation~\ref{enoise}) for each channel $i$ to the standard normal, for all 160 channels. For p-values above 0.05 we cannot reject the null at the 5\% significance level. The dotted line shows the distribution of the K-S test probabilities for 160 sets of random points drawn from a $\mathcal{N}(0,1)$ distribution. Right: Q-Q plot of noise quantiles vs. standard normal quantiles, for a set of 16 channels, color-coded from the lowest (indigo) to the highest (red) frequency. The corresponding K-S p-values are also color-coded. If the two distributions are identical, the points should lie along the diagonal. Overplotted with a dotted line is a simulated observation, with points randomly drawn from a $\mathcal{N}(0,1)$ distribution, showing a scatter similar to our channels. The corresponding p-value is shown in black in the top left corner.}
\label{gnoise} 
\end{figure}

To be able to simulate the estimator behavior in the absence of any signal, we need to start by choosing a noise distribution. For our simulations, we assume that the noise is gaussian distributed, with a standard deviation given by the measured error in each channel. To test this assumption, we study the noise in a 6.5 hour Z-Spec integration on blank sky, recorded on May 11 and 12, 2010. For each channel $i$ for our blank-sky data, we look at the quantity

\begin{equation}\label{enoise}
 G_{ik}=\frac{S_{ik}-A_{i}}{\sigma_{ik}},
 \end{equation}
 
\noindent where $k$ represents the nod number, $i$ is the channel number, $A_{i}$ is the average signal in channel $i$ over all nods, and $S_{ik}$ and $\sigma_{ik}$ are the signal and the noise, respectively, for the corresponding nod and channel. In our blank-sky data there are about 220 nods per channel, after flagging. The distribution of $G_{ik}$ for a given channel $i$, over all nods, should be a standard normal if our assumptions are correct. 

We first apply a Kolmogorov-Smirnov test \citep[K-S, ][]{Kol,Smir}, which tests the hypothesis that the observed noise distribution is drawn from a standard normal by comparing their cumulative distributions. The test results are quantified in terms of the p-value, which is the probability that a value of the test statistic equal or greater than the one observed would be obtained if the null hypothesis were true. In the left panel of Figure~\ref{gnoise}, we show a histogram of the p-values of the K-S statistic for all the channels, which demonstrates that we cannot reject the null hypothesis that the noise is gaussian distributed at the 5\% level for any of the channels, since all p-values lie above this level. The plot also shows a large spread in p-values from channel to channel. For comparison, we run the same K-S test for 160 sets of random numbers drawn from a $\mathcal{N}(0,1)$ distribution. Each set contains the same number of samples as the corresponding channel. The distribution of the K-S p-values for these computer-generated normal samples is shown by the dotted line in the left panel of Figure~\ref{gnoise}. This simulation also shows a large spread in p-values, approximately uniform across the range.

The K-S statistic can be affected by a variety of factors, pertaining both to departures from gaussianity (shape of the distribution), and to mismatches in the parameters of the assumed distribution (i.e., the sample is drawn from $\mathcal{N}(\mu_0,\sigma_0^2)$ instead of $\mathcal{N}(0,1)$). If our assumption that the noise for each channel is gaussian distributed is correct, but we have over- or under-estimated $\sigma_{ik}$, this can, in principle,  result in small p-values
for the K-S test. A useful tool in this case is the quantile-quantile (Q-Q) plot \citep{qq}, which is more sensitive to multiple aspects of the distributions being compared, but does not provide a quantitative measure of these deviations. 

A Q-Q plot is basically a representation of the observed data quantiles versus the theoretical quantiles of the assumed distribution ($\mathcal{N}(0,1)$ in this case). The quantiles are defined as regular intervals on the cumulative distribution function, intuitively intervals of equal probability. The Q-Q plot is demonstrated for a sample of 16 channels in the right panel of Figure~\ref{gnoise}. If the noise is gaussian distributed {\it and} the $\sigma_{ik}$ are estimated correctly, the points on Figure~\ref{gnoise} for each channel will follow the diagonal. If the noise is over- or under-estimated, the relationship will be still linear, but with a different slope. Figure~\ref{gnoise} shows that this might be the case for some of the low-frequency channels (blue), for which the noise is known to be more variable due to both intrinsic bolometer problems and atmospheric noise. These very-low frequency channels are in fact excluded from our redshift-finding algorithm.

Departures from gaussianity, which could be due to noise correlations between channels, would stand out in the Q-Q plot as departures from linearity. For comparison, we overplot in Figure~\ref{gnoise} the curve obtained for a computer-generated sample drawn from $\mathcal{N}(0,1)$ (black dotted line),  which shows a similar level of scatter around the diagonal as our channels. We conclude that the scatter of the noise distribution around the linear correlation in the Q-Q plot is negligible when compared to the results from the random number generator for a standard normal, supporting the results of the K-S test for the gaussianity of the channel noise. 

Our simulations create multiple realizations of a pure-noise spectrum, with the signal in each channel being a gaussian random variable with mean 0 and standard deviation equal to the measured noise in that channel, $\mathcal{N}(0,\sigma_i^2)$. Even if for some channels the noise might be over- or under-estimated, its exact value is not essential, since it cancels out as part of the signal-to-noise ratios in estimator definitions, and in the end we are left with $\mathcal{N}(0,1)$ distributions for the estimators (see Appendix). As discussed above, the noise per channel from our blank-sky observation is well approximated by a gaussian distribution, aside from small channel-to-channel noise correlations. In our simulations, we reproduce these residual noise correlations between different channels using the method of Cholesky factorization. The noise correlation matrix is constructed from all the nods contained in the blank sky spectrum,

\begin{equation}\label{ecovar}
 C_{ij}=\frac{\sum_{k}(S_{ik}-A_{i})(S_{jk}-A_{j})}{sqrt{\sum_{k}(S_{ik}-A_{i})^2\sum_{k}(S_{jk}-A_{j})^2}},
 \end{equation}

\noindent where the sums are taken over all nods. After multiplying a randomly-generated uncorrelated vector with the lower-triangular matrix from the Cholesky decomposition, one obtains a vector with the same correlation properties as the original sky noise model \citep[e.g., ]{Kaiser1962}.

\begin{figure*}
\begin{center}
\includegraphics[width=2.7in,angle=90]{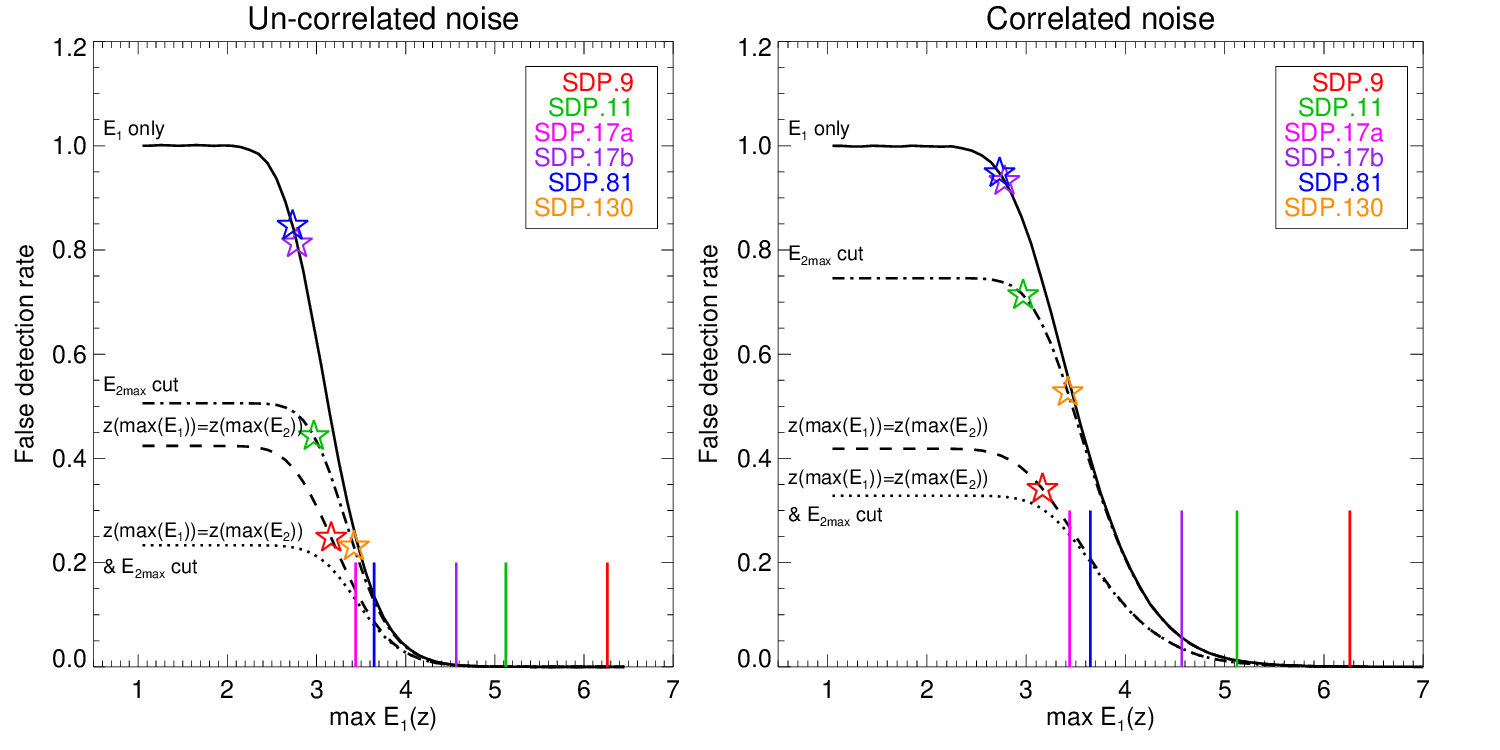}
\end{center}
\vspace{-0.2in}
\caption{$FDR$ distributions resulting from noise simulations with un-correlated and correlated noise (left and right, respectively). The curves show the decrease in $FDR$ as a function of the $E_1$ value, before (solid) and after applying the condition $E_{2max}\geq2.12$ (dash-dot), $z(max(E_1))=z(max(E_2))$ (dashed), and both conditions jointly (dotted). The solid vertical lines correspond to the $E_1$ values at the determined redshift for each target. These redshifts satisfy both conditions (dotted line), and the corresponding $FDR$ for each of them is better shown in Figure~\ref{fig:sig}. For SDP.130 the values of the estimators at z=2.626 (GBT/Zpectrometer) are negative, and therefore cannot be maxima (no corresponding vertical line). The stars show the $FDR$ associated with the values of $max(E_1)$ obtained after subtracting the lines from each spectrum (the spectrum of SDP.130 is used as-is, and for SDP.17 after subtracting both sets of lines), each of them being placed on the curve corresponding to the conditions satisfied by the residual maximum. After subtracting the lines, none of the maxima satisfies all our criteria for redshift determination, and the individual significance for each of them can be read off these curves.}
\label{fig:fdr} 
\end{figure*}

\begin{figure*}
\begin{center}
\begin{tabular}{cc}
\hspace{-0.5in}
\includegraphics[width=2.7in,angle=90]{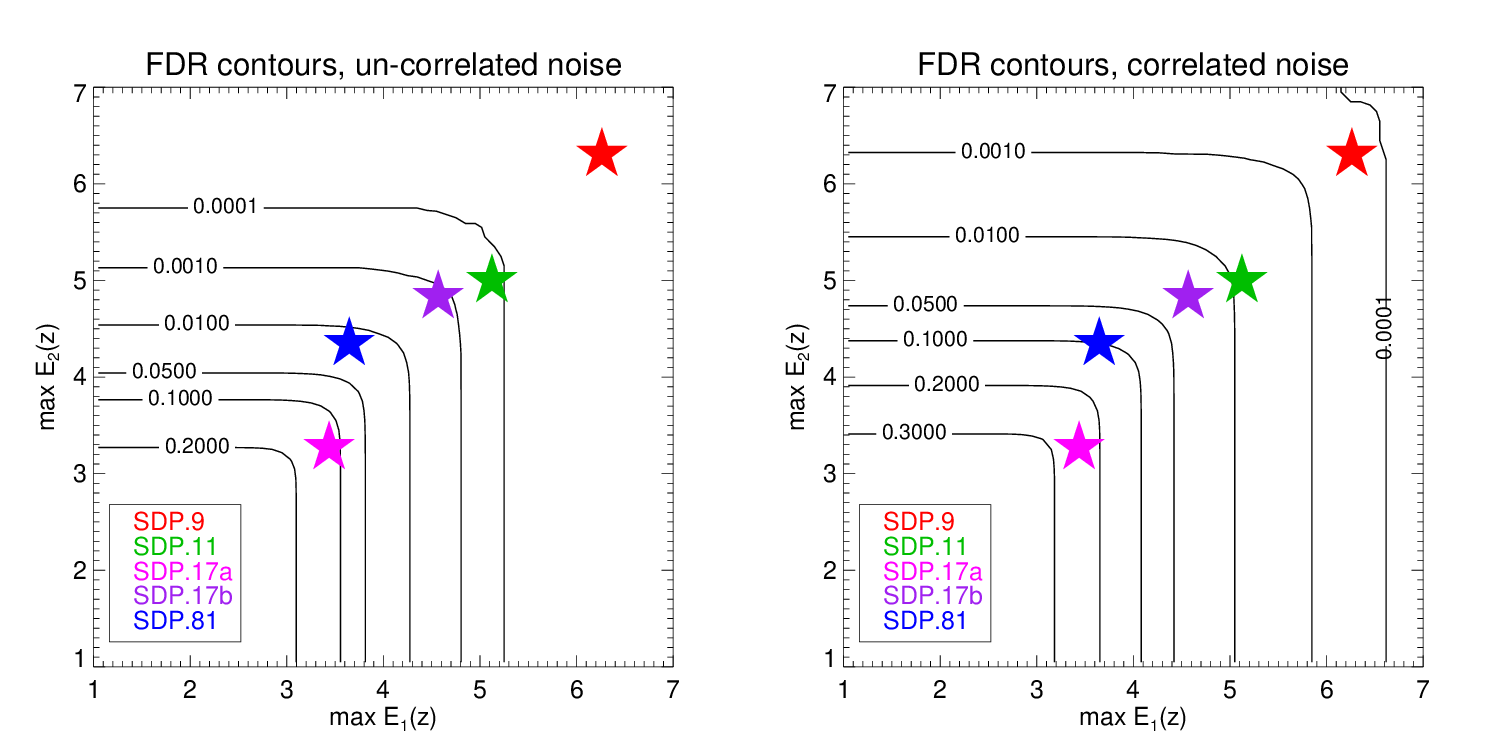}
\end{tabular}
\end{center}
\vspace{-0.2in}
\caption{$FDR$ contour plots from our simulations, as a function of $E_1$ and $E_2$ values. The points corresponding to the maxima of $E_1$ and $E_2$ for each of the sources that pass the cuts (all except SDP.130, see Figure~\ref{A981zearch}) are shown by the filled stars. The contours correspond to the dotted line in Figure~\ref{fig:fdr} (satisfying both $z(max(E_1))=z(max(E_2))$ and $E_{2max}\geq2.12$ conditions). The points obtained after line subtraction do not satisfy either condition and therefore are not shown.}
\label{fig:sig} 
\end{figure*}

We run separate simulations for correlated and un-correlated noise, each with $10^5$ realizations of noise spectra. For each realization of the noise spectrum, we record the maximum values of the estimators $E_1$ and $E_2$, and construct their joint distribution function over all realizations. For any $measured$ $max(E_1)$ and $max(E_2)$ from real data, we define the associated false detection rate ($FDR$) as the probability of finding a maximum value of $E_1>max(E_1)$ and $E_2>max(E_2)$ by chance, in the absence of real signal. We calculate this joint probability from the simulated 2-D right-cumulative distribution function of the maxima of the two estimators, as shown in Figure~\ref{fig:sig} for each measured $(max(E_1),max(E_2))$ pair. Figure~\ref{fig:fdr} shows the marginal $FDR$ (the 2-D cumulative distribution marginalized over $E_2$) for all the $max(E_1)$ values (solid black line), as well as for the $max(E_1)$ values left after imposing the additional constraints on the estimators discussed below. 

As the first constraint, for each simulated spectrum we identify all the $max(E_1)$ and $max(E_2)$ values that satisfy the condition that both estimators reach their maxima at the same redshift. As can be seen from the dashed line in Figure~\ref{fig:fdr}, imposing this condition reduces the total $FDR$ to about 40\% for both correlated and un-correlated noise. The decrease in $FDR$ is due to the fact that the {\it locations of the maxima} of $E_1$ and $E_2$ deviate from a perfect correlation (Pearson correlation coefficient $<0.80$), with the scatter spread over all redshifts. This shows that the combination of two test statistics is more robust against random fluctuations than any estimator used independently, and considerably reduces the noise floor across the redshift range.

Requiring that the two estimators be maximized at the same redshift we still get a rather high total $FDR$ (at least 40\%). In order to further reduce the number of spurious redshifts obtained from blank sky spectra, we introduce a signal-to-noise threshold cut. We define the quantity 

\begin{equation}\label{e2max}
\begin{split}
 E_{2max}(z)=\mathrm{max}\{& f_{ij} | f_{ij}=0.5(S_{i}/\sigma_{i}+S_{j}/\sigma_{j}),\\
 & 1\leq i,j\leq N(z), i < j\}.
 \end{split}
 \end{equation}

\noindent This definition is very similar to $E_2(z)$, with the $median$ replaced by the $max$, and without the normalization factor. The normalization factor is not needed here because we want to be able to establish a threshold criterion across the entire redshift range, independent of the number of lines $N(z)$. Since this is not an estimator, we are not interested in standardizing its distribution, and moreover, its distribution will be likely different from that of $E_2$, as an extreme order statistic rather than a central order statistic.

\begin {figure}[h]
\centering
\includegraphics*[scale=0.5,angle=0]{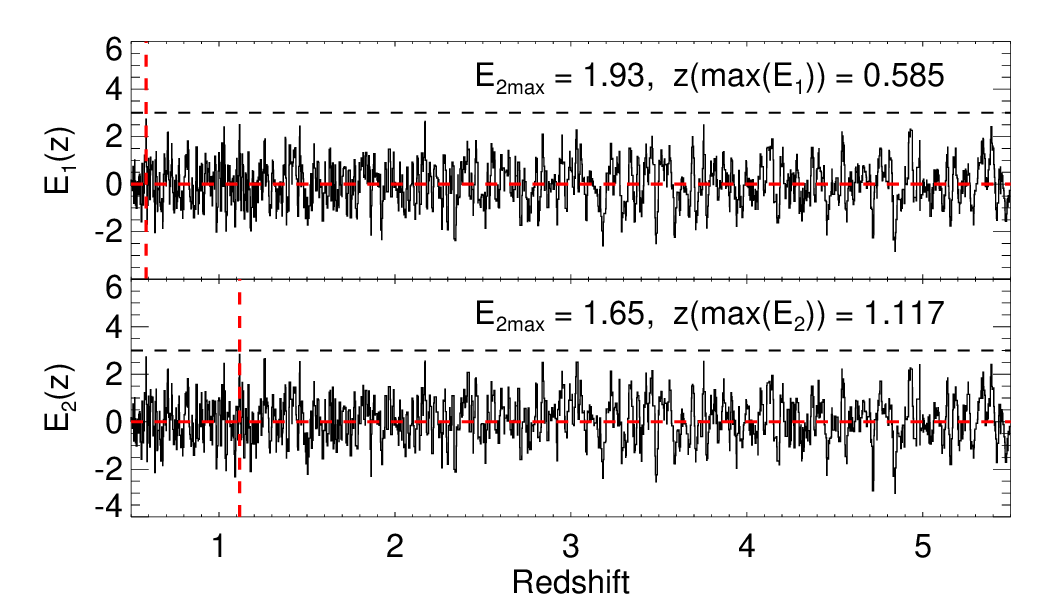}
\caption{Null test for the redshift finding algorithm, using a blank sky spectrum. The dashed horizontal lines are drawn at 0 and 3, to guide the eye. Note that the values of the estimators are below 3 across the redshift range, the positions of their maxima do not coincide, and the $E_{2max}$ values corresponding to these maxima are below our threshold. The dashed vertical lines indicate the positions z(max(E$_{1}$), and z(max(E$_{2}$)), respectively. \label{skyzearch}}
\end{figure}

For any $i$ and $j$ the quantity $f_{ij}=0.5({S_{i}}/{\sigma_{i}}+{S_{j}}/{\sigma_{j}})$ is distributed as a normal with standard deviation $\sigma_{av}=1/\sqrt{2}$ (since $S_{i}/\sigma_{i}$ has a standard normal distribution).  We choose the value $E_{2max}(z)=3\sigma_{av}=2.12$, as our threshold. In other words, we require that strongest pair of lines at the determined redshift have an average signal-to-noise greater than 2.12. This places approximate limits on the signal-to-noise per channel and the values of the estimators for the determined redshift of $\sim$2 and $\sim$3, respectively. If no signal is present in the spectrum (null hypothesis), the distributions of $E_1(z)$ and $E_2(z)$ are well approximated by standard normals for all redshifts $z$ (see Figure~\ref{estdist}). As such, our noise cut also implies rejecting the null hypothesis at the $\sim$99\% level for the determined redshift. A realization of the two test statistics using the blank sky spectrum is shown in Figure~\ref{skyzearch}. In this case, the maxima of the two statistics occur for different redshifts, the values of $E_{2max}$ are below our threshold for both $E_1$ and $E_2$ maxima, and the values of both estimators are below 3 for all redshifts.

The effect of imposing the $E_{2max}$ constraint on the $FDR$ is shown by the dot-dashed curve in Figure~\ref{fig:fdr}. The noise threshold criterion improves the significance (1-$FDR$) of the lowest signal-to-noise results by cutting the limiting $FDR$ down to about 50\% for un-correlated noise and to $\sim75\%$ for correlated noise. The joint effect of both constraints is shown by the dotted curve in Figure~\ref{fig:fdr}, and the corresponding 2-D $FDR$ distributions as a function of the maxima of both $E_1$ and $E_2$ are plotted in Figure~\ref{fig:sig}. We derive that the estimator values passing these 2 tests will result in a total false detection rate lower than 24\% for uncorrelated noise and below 33\% for correlated noise. 

\begin{deluxetable*}{cccccccccc}
\tabletypesize{\footnotesize}
\tablewidth{0pt}
\tablecaption{\label{tab:cont}Summary of the H-ATLAS galaxy sample and the parameters derived from fitting their submm SEDs.}

\tablehead{
{H-ATLAS}& {$\mu$} & {$z$} & {Significance}  & {$\mu$~$L_{IR}$} & {$T_{d}$\tablenotemark{(d)}} &{$\alpha$}&  {$\mu$~$M_{d,lim}$\tablenotemark{(e)}} & {$\mu\Omega_{d}$\tablenotemark{(e)}} & {$\mu$~$SFR$}  \\
{SDP ID} & & & {(\%)} &{(10$^{13}$ L$_{\odot}$)}&{(K)} & & (10$^{9}$ M$_{\odot}$) & (arcsec$^{2}$) &(10$^{3}$ M$_{\odot}$/yr) }
\startdata
 SDP.9 & ... & 1.577$\pm$0.008 & 100 (99.97) & 4.4$\pm$0.5 & 57$\pm$1 & 3.8$\pm$0.2 & 2.5& 0.65 & 6.6$\pm$0.8\\
 SDP.11  & ... & 1.786$\pm$0.005 & 99.98 (99.22) & 7.8$\pm$0.9 & 69$\pm$1 & 5.7$\pm$0.4 & 1.7 & 0.43 & 11.7$\pm$1.3\\
 SDP.17a\tablenotemark{(a)}  & ... & 0.942$\pm$0.004 & 87.33 (74.77) & 0.4$\pm$0.09 & 27$\pm$1 & 2.9$\pm$0.1 & 4.9 & 1.44 & 0.6$\pm$0.1\\
 SDP.17b\tablenotemark{(a)}  & ... & 2.308$\pm$0.011 & 99.86 (97.46) & 3.9$\pm$0.9 & 66$\pm$1 & 2.9$\pm$0.1 & 1.1 & 0.30 & 5.8$\pm$1.5\\
 SDP.81  & 18-31\tablenotemark{(b)} & 3.037$\pm$0.010 & 98.26 (90.02) & 6.4$\pm$0.3 & 58$\pm$1 & 3.2$\pm$0.1 & 2.2 & 0.69 & 9.6$\pm$0.4\\
SDP.130  & 5-7\tablenotemark{(b)} & 2.626$\pm$0.0003\tablenotemark{(c)} & N/A & 4.3$\pm$0.2 & 55$\pm$1 & 2.7$\pm$0.3 & 1.6 & 0.47 & 6.5$\pm$0.3\\

\enddata
\tablecomments{The columns list: 1) the ID of the source in the SDP H-ATLAS catalogue; 2) the gravitational lensing magnification factor; 3) the measured redshift; 4) the redshift significance, calculated as $1-FDR$, where the $FDR$ has been defined in Section~\ref{noise}. The significance for correlated noise is given in parenthesis; 5) the integrated IR luminosity, obtained as the average between the SED fits with the CE01 libraries and the DH02 libraries. The factor $\mu$ is shown in front of quantities affected by gravitational lensing magnification; 6) the dust temperature, with the caveats described in the text; 7) the index of the power-law continuum fit to Z-Spec data; 8) the dust mass; 9) the solid angle subtended by the dust emitting region; and 10) the star formation rate. }
\tablenotetext{(a)} {The total observed flux was split between the two components, using a frequency-independent scale factor.}
\tablenotetext{(b)} {Values taken from \citet{Negrello10}.}
\tablenotetext{(c)} {Redshift determined by GBT/Zpectrometer, followed by a more precise measurement with PdBI/IRAM \citep{Negrello10}.}
\tablenotetext{(d)} {The uncertainties for $T_{d}$ are likely underestimated. The values shown are formal errors from the fit, and don't include correlations between parameters, or the inaccuracy of the assumed shape of the SED model. The values for $\beta$ and $\nu_{0}$ are kept fixed for all sources.}
\tablenotetext{(e)} {Calculated in the optically thin limit. The dust masses should be interpreted as robust lower limits for the true total dust mass in the galaxy (see text).}

\end{deluxetable*}

We emphasize that these $FDR$ values are {\it limiting values}, in the sense that they are $independent$ of the actual values of $max(E_1)$ and $max(E_2)$, and only satisfy the requirement that the estimators pass the two tests (attain maxima at the same redshift and exceed the $E_{2max}$ threshold). For any actual redshift determination, the associated false detection rate will be determined by the values of $max(E_{1})$ and $max(E_{2})$ for that particular spectrum, which are inversely correlated with the $FDR$, as indicated by the vertical lines in Figure~\ref{fig:fdr}. For the primary redshifts determined in this paper, we find that the integrated false detection rates are smaller than 2\%, as described in Section~\ref{iz}, and listed in Table~\ref{tab:cont}. Note that all quoted $FDRs$ are by definition {\it integrated over the whole redshift range}, and represent the probability of obtaining a false positive, at any redshift, given the values of the estimator maxima, but the probability of obtaining {\it the same} redshift by chance is roughly a factor of $\sim$5000 lower, based on the number of redshift bins searched.  This is because it is even more unlikely that the same channels fluctuate high due to noise alone.

To summarize, a redshift $z_{0}$ is accepted when the following conditions hold:
 
 \begin{equation}\label{acceq}
 \begin{aligned}
 E_{1}(z_{0})=max(E_{1}),\\
 E_{2}(z_{0})=max(E_{2}),\\
 E_{2max}(z_{0})\geq 2.12,
 \end{aligned}
 \end{equation}

\noindent where the last condition is basically a signal-to-noise threshold criterion, and the significance of the estimated redshift is calculated as 1-$FDR$, with the $FDR$ derived from the noise simulations, as explained above.
  
At the CSO, Z-Spec can reach a measured maximum sensitivity of 0.5~Jy~s$^{1/2}$ per channel for an atmospheric optical depth $\tau_{225}$=0.068 \citep{inami08}. Combining the signal-to-noise threshold criterion with the measured sensitivity of Z-Spec, we estimate that a redshift can be determined in less than 1.4 hours of integration time if the line flux densities per channel are on the order of 15~mJy, but can require more than 12.6 hours if the flux density is less than 5~mJy. For our galaxy sample, the mean integrated CO line flux (Table~\ref{tab:lco}) is $\sim$18~Jy~km~s$^{-1}$, while the average width of the channels is 950~km~s$^{-1}$. However, the flux density per channel could be only $\sim$10~mJy if the line flux happens to be split between two adjacent channels. In this case, the typical integration time for obtaining a redshift with Z-Spec would be at least 3.5 hours. These time estimates reflect closely the best performance of the instrument and do not include calibration overheads. The actual integration time needed to obtain a redshift will depend strongly on the instrument sensitivity at the time of the observations.

 \subsection{Redshifts for the H-ATLAS SDP Sample }\label{zdet}
 
\begin {figure*}
\centering
\includegraphics*[scale=0.55,angle=0]{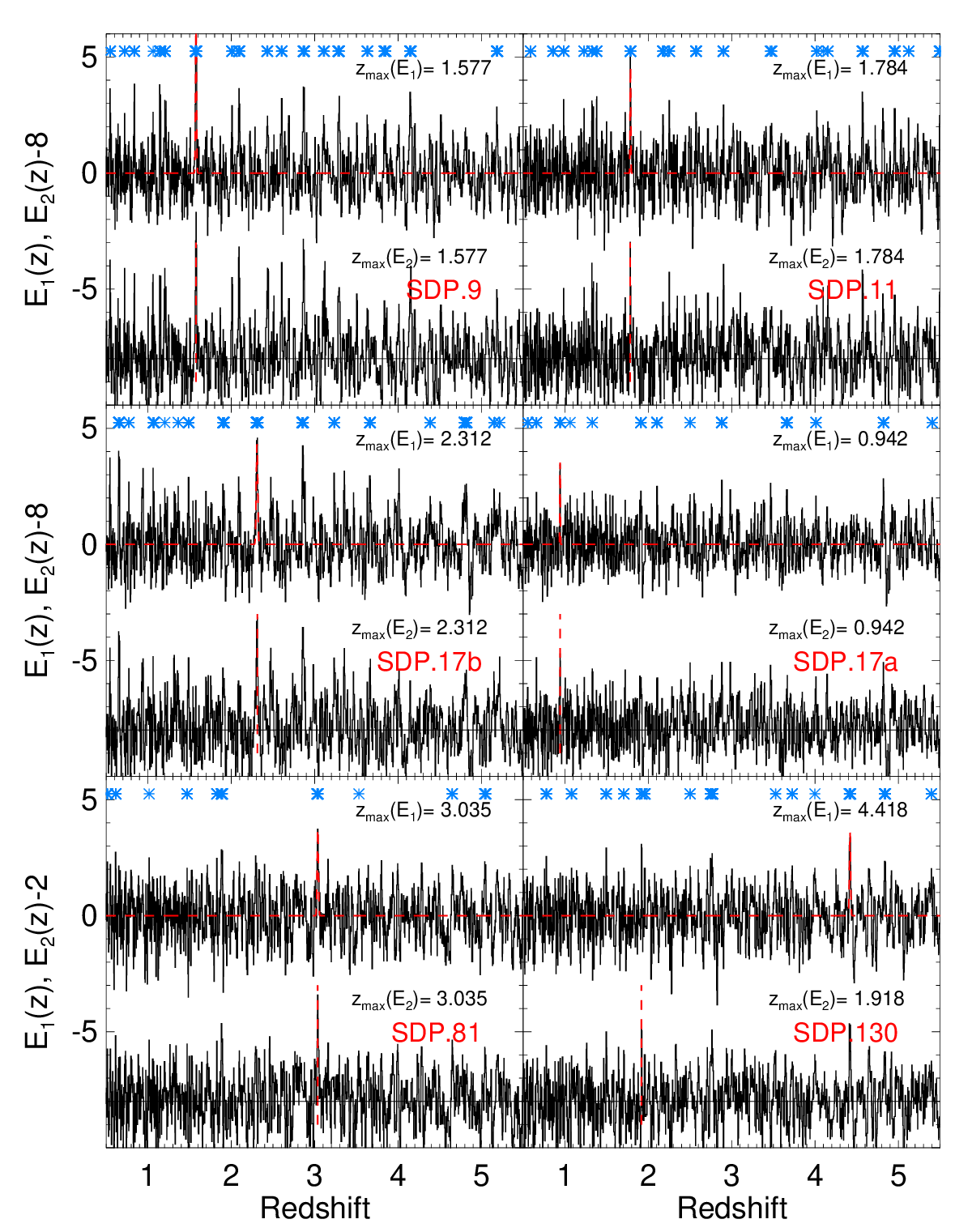}
\caption{Results of running the redshift-finding algorithm for all the H-ATLAS sources in our sample. The $E_{2}$ test statistic has been offset vertically by 8 units, for clarity. The blue asterisks show the positions of the largest secondary peaks arising from coincidences with the lines from the actual redshift (see text). These peaks contain the same information as the main peak. In the SDP.17 panel, we note the extra peaks that do not match the secondary peaks corresponding to the first selected redshift. The SDP.17a panel shows the determination of the second redshift from the same spectrum, after subtracting the high-redshift component. No redshift is determined for SDP.130. \label{A981zearch}}
\end{figure*}

The results of applying this algorithm to our galaxy sample is shown in Figures~\ref{A981zearch} and \ref{fig:sig}. We secure the redshifts for four out of five sources, with an FDR $<10$\% in all cases. The redshift value and its uncertainty (Table~\ref{tab:cont}) are determined from the position and width of the peak of the $E_{1}$ test statistic (Figures~\ref{A981zearch}). Due to the finite width of the spectral channels, nearby redshifts can have the same or similar significance, since the lines will fall on the same channels for a narrow range of redshifts, given by our redshift space sampling. As we go further from the real redshift, some of the lines might still fall on the same channels, but not all of them, so the value of E$_{1}$ will drop. We fit a gaussian to the peak of $E_{1}(z)$, and define the redshift error as the upper limit for the standard deviation of this gaussian. This value is at least as large as the channel width, and accounts for the varying channel widths across the bandpass. 
 
For all the galaxies except SDP.130, the maxima of $E_{1}$ and $E_{2}$ satisfy both our criteria for a secure redshift determination, and for SDP.17 we can identify a second redshift satisfying our criteria after subtracting the first set of lines from the spectrum. We calculate the significance of the redshift for each of our sources by interpolating the $FDR$ at the observed values of $max(E_1)$ and $max(E_2)$. Figures~\ref{fig:sig} and \ref{fig:fdr} show the derived estimator values for each source relative to the $FDR$ distribution, and the significance of each redshift determination can be read off directly from these figures. Of all redshifts that passed our criteria, the redshift of SDP.17a (the second derived for SDP.17) has the lowest significance, close to 90\% (75\% for correlated noise), while all primary redshifts have a significance of at least $\sim99\%$ (90\% for correlated noise), equivalent to $\sim3\sigma$ or greater for a gaussian distribution.

When lines are present in the spectrum, aside from the main peak due to the true redshift, secondary peaks will arise in the $E_{1}$ and $E_{2}$ distributions, corresponding to redshifts where some of the lines in the line list fall on the same channels as the observed lines. The secondary peaks are marked by blue asterisks for each source in Figure~\ref{A981zearch}. The real redshift will have higher significance than the redshifts corresponding to these secondary peaks, since the largest number of lines add their contribution to the total signal in this case. 

After removing the lines corresponding to the measured redshifts from the spectra (both redshifts for SDP.17), the secondary peaks in the $E_{1}$ and $E_{2}$ distributions are reduced to the noise level, and the maxima of $E_1$ and $E_2$ fail to satisfy one or or both of our criteria. The $FDRs$ associated with these line-subtracted spectra are plotted as stars in Figure~\ref{fig:fdr}, indicating the significance of the remaining features. The stars are vertically positioned on the false detection curve corresponding to the criteria satisfied by $max(E_{1})$ after removing the lines. The computed FDRs and the fact that they do not satisfy both criteria indicate that in all cases the results after line subtraction are consistent with noise.

All the measured redshifts, with their error bars and associated significance (calculated as $1-FDR$) for both correlated and un-correlated noise, are listed in Table~\ref{tab:cont}. The significance of our redshift determinations together with the statistical redshift determination criteria (Eq.~\ref{acceq}) show that in general a redshift was already secured by Z-Spec after an integration time much shorter than the total integration time listed in Table~\ref{tab:obs}. Future submm instruments with better sensitivity will be able to obtain the redshifts of such galaxies even faster, and open the possibility of large submm redshift surveys.

 \subsection{Comments on Individual Redshifts }\label{iz}
 
The individual redshifts are presented in the order of the observations (see Table~\ref{tab:obs}).

\begin{deluxetable*}{ccccccc}
\tabletypesize{\footnotesize}
\tablecaption{\label{tab:lco} Integrated fluxes for the emission lines identified in each galaxy.}

\tablehead{{Line} & {Frequency}& & & Integrated Line Flux (Jy km s$^{-1}$)&&\\
&(GHz)&{SDP.9}&{SDP.11}&{SDP.17a} &{SDP.17b} & {SDP.81} }

\startdata
CO (4$-$3)   				& 461.041 & ... 				& ... 				& 14$\pm$9 	&...				& ...  			\\
CO (5$-$4)  		 		& 576.268 & 25$\pm$5 	& 23$\pm$8 	& 27$\pm$9\tablenotemark{(b)}	&...			& ...  			\\
CO (6$-$5)   				& 691.473 & 33$\pm$7 	& 29$\pm$10 	& ... 				&17$\pm$5	& ...  			\\
CO (7$-$6)   				& 806.652 & ... 				& 18$\pm$14\tablenotemark{(a)}	& ... 				&11$\pm$7\tablenotemark{(a)}	& 12$\pm$4\tablenotemark{(a)} \\
CO (8$-$7)   				& 921.800 & ... 				& ... 				& ... 				&16$\pm$6	&  5$\pm$3 \\
CO (9$-$8)     			& 1036.91 & ... 				& ... 				& ... 				&...				& 6$\pm$3  \\
CO (10$-$9)   			& 1151.99 & ... 				& ... 				& ... 				&...			& $<$~6.5 \\\relax
[\ion{C}{1}] $^{3}P_{1}\rightarrow^{3}$$P_{0}$   	& 492.160 & $<$~11 	& ... 				& $<$~6 	&...			& ...  			\\\relax
[\ion{C}{1}] $^{3}P_{2}\rightarrow^{3}$$P_{1}$	& 809.342 & ... 				& 31$\pm$14\tablenotemark{(a)} 	& ... 				&13$\pm$7\tablenotemark{(a)}	& $<$~6.5\tablenotemark{(a)} \\
H$_{2}$O $2_{0,2}-1_{1,1}$& 987.914 & ... 				& ... 				& ... 				&19$\pm$7\tablenotemark{(b)}	& ... 			\\

\enddata
\tablecomments{The columns list: 1) the transition; 2) the rest frame frequency of the transition; and 3) the integrated line flux for each galaxy (as measured, uncorrected for dust absorption) with 68\% confidence intervals. Upper limits are 3$\sigma$.}
\tablenotetext{(a)} {These lines originate in the same source and are blended at the Z-Spec resolution. The error bars account for this uncertainty.}
\tablenotetext{(b)} {In the spectrum of SDP.17, the water line at z=2.308 and the CO(5$-$4) line at z=0.94 are blended.}
\end{deluxetable*}
 
{\it SDP.81} The redshift for SDP.81, z=3.037$\pm$0.01, obtained by this method on 19 March 2010, was confirmed (z=3.042$\pm$0.001) with follow-up observations with the IRAM Plateau de Bure Interferometer on 23 March 2010 \citep[IRAM/PdBI, ][]{Negrello10,Neri10} and with an independent blind search on 25 March 2010 by the Zpectrometer instrument at the Green Bank Telescope \citep[GBT/Zpectrometer, ][]{Negrello10,Frayer10}. Both follow-ups were informed by a concurrent photometric redshift estimate (2.9$^{+0.2}_{-0.3}$). With the possible exception of the second redshift for SDP.17, this is the redshift with the lowest significance in our sample, with an $E_1$ peak of 3.8, due to the weakness of the CO lines beyond CO($7-6$). This is the first blind redshift obtained by Z-Spec.

{\it SDP.130} SDP.130 has a redshift of 2.6260$\pm$0.0003, measured by GBT/Zpectrometer \citep[z=2.625$\pm$0.001, ][]{Frayer10}, and made more precise with PdBI/IRAM \citep{Negrello10,Neri10}. So far, three CO lines have been measured in this galaxy at this redshift, namely the CO(1$-$0) line observed with the Zpectrometer, and the CO(3$-$2) and CO(5$-$4) lines observed with PdBI \citep{Negrello10}, on a tuning that was successfully guided by the sub-mm photometric redshift of $z=2.6^{+0.4}_{-0.2}$ \citep{Negrello10}. However, we do not detect any of the higher $J$ transitions ($J_{u}>$6) that would fall in the Z-Spec bandpass at this redshift. The values of our estimators for z=2.625 are negative, suggesting that there is no signal left in the spectrum at this redshift after continuum subtraction. The estimators do not pass our redshift determination criteria for any other redshift, and Figure~\ref{fig:fdr} shows the significance of the maximum $E_1$ value obtained under these conditions (orange star). 
This non-detection, which places upper limits on the integrated fluxes of the CO(6$-$5) through (9-8) lines of $<$~12.5~Jy km s$^{-1}$, suggests a low ($<$~50 K) gas temperature in the z=2.626 galaxy. We attempted to identify the line at 277~GHz, marked in red in Figure~\ref{spectra}, with the CO(3$-$2) transition at z=0.25, but that would be inconsistent with the optical spectroscopic redshift of the lensing galaxy \citep[0.220$\pm$0.002, ][]{Negrello10} by more than 7000~km~s$^{-1}$, as well as inconsistent with the observed SED. Based on the correlation beween the CO and the total infrared luminosity, the observed luminosity of the CO(3$-$2) line would correspond to an ULIRG-class object at z=0.25, which would dominate the SED at 250~$\mu$m. No separate 250~$\mu$m-bright object is found nearby, and the PACS and SPIRE photometry of SDP.130 (see also Section~\ref{sseds}) is inconsistent with the two sources being blended. Similarly, identifying this feature with the 987~GHz water line at z=2.626 would require a velocity offset of $\sim$4200~km~s$^{-1}$, and usually the presence of highly excited CO gas, which is not observed. This feature remains unidentified.

{\it SDP.17} Given the size of the Z-Spec beam (FWHM$\approx$30\arcsec) and the possible presence of lensing or other foreground structures in the same beam, the observed spectrum could be a combination of features from multiple objects. We choose this interpretation for the spectrum of SDP.17, best described by two components at different redshifts (both listed in Table~\ref{tab:cont}). The first redshift found by our algorithm is 2.308 (SDP.17b). After fitting the CO lines at this redshift and subtracting them from the spectrum, we perform a second redshift determination, identifying a second component with a redshift of 0.942 (SDP.17a). This combination explains all the features present in the spectrum (see Figure~\ref{spectra17}), and is consistent with the interpretation of the 299~GHz feature as the restframe 987~GHz water line at a redshift of 2.308. This water line has been seen to be very strong in other AGN and star-forming galaxies at low redshift, such as Mrk231 and Arp 220 \citep{Gonzales2010}, and it has been tentatively detected in the Cloverleaf quasar at z=2.56 by \citet{bradford09}. More recently, multiple excited water trasitions have also been detected in the quasar APM 08279+5255 at z=3.91 \citep{Bradford2011,Lis2011,vdWerf2011}. The redshifts of SDP.9 and SDP.17b have recently been confirmed by follow-up observations of the CO(2$-$1) and (3$-$2) lines, respectively (L. Leeuw, private communication), with CARMA. Recent IRAM/PdBI observations \citep{Omont2011} have confirmed the water line at z=2.3052, but did not find any other high significance line in the bandpass, which does not exclude the possibility of a CO(5$-$4) line at z$>0.944$. The second redshift (SDP.17a) has a much lower significance, but it is in agreement with the photometric and spectroscopic optical redshifts \citep[0.77$\pm$0.13 and 0.9435$\pm$0.0009, respectively][]{Negrello10}. Alternatively, the peak now identified with the CO(5$-$4) line at z=0.94 could be arising from correlated noise fluctuations with the nearby water line. To confirm the presence of CO at z=0.94, we are planning a follow-up of the CO(4$-$3) line. The presence of multiple ULIRGs in a single line of sight is intriguing, and is an example of discoveries that can be made possible by Z-Spec's broad bandwidth. It also raises the possibility that the flux-limited sample is affected by chance alignments, and the presence of multiple sources in the beam. However, this is likely a negligible effect for lensed sources, as the continuum sub-mm and mm flux will be clearly dominated by the lensed, high-z galaxy, and not by the foreground lens \citep{Negrello10}.
 
{\it SDP.9 and SDP.11} The significance of the redshifts for these galaxies corresponds to $max(E_1)$ values of 6.5 and 5.3, respectively (Figures~\ref{fig:fdr} and {fig:sig}). The redshift of SDP.9 has been confirmed by CARMA observations, and more follow-up observations are currently planned for both SDP.9 and SDP.11.

 \section{GAS AND DUST PROPERTIES}\label{sgalaxy}

A model including the lines and power-law continuum is fit to each spectrum in Figures~\ref{spectra} and \ref{spectra17}, allowing the line intensities, redshift, and continuum slope to vary. The best fit power-law index $\alpha$ for each galaxy is listed in Table~\ref{tab:cont}. The initial estimate for the redshift is provided by the algorithm described above, and the fit is constrained by the requirement that all the lines be at the same redshift. In cases where some of the lines are blended, we first fit only the unblended lines to obtain a more precise value for the redshift, and then we fit all the lines simultaneously, with the redshift kept fixed, to get the integrated line strengths, listed in Table~\ref{tab:lco}. Although the lines are not resolved, the signal from one line can be spread among adjacent channels due to the overlap of their frequency responses. We measure only the integrated line strengths, taking into account the frequency response of each Z-Spec channel, weighted according to the line width. On average, line widths below $\sim$1000~km~s$^{-1}$ are not resolved by Z-Spec, and we choose a value of 300~km~s$^{-1}$ in fitting the integrated line strengths. This value closely matches the width of the lines for SDP.81 and SDP.130 at PdBI \citep{Neri10}, but is relatively low compared to the range found by interferometric measurements of other lensed high-redshift galaxies \citep{greve05,Knudsen2009}. The CO(1$-$0) line widths determined by the GBT/Zpectrometer are somewhat larger ( 435$\pm$54~km~s$^{-1}$ for SDP.81 and 377$\pm$62~km~s$^{-1}$ for SDP.130), suggesting that an additional gas component might contribute to this line. However, the determination of the integrated line fluxes is not sensitive to the choice of the line width up to values of the order of the channel width. The largest uncertainties in the integrated line strengths arise in the case of line blending, such as the CO(7$-$6) and [\ion{C}{1}] $^{3}P_{2}\rightarrow^{3}$$P_{1}$ lines, or the overlapping lines at different redshifts in SDP.17 (blended lines are indicated in Table~\ref{tab:lco}).  

\begin {figure*}
\centering
\includegraphics[scale=0.65,angle=0]{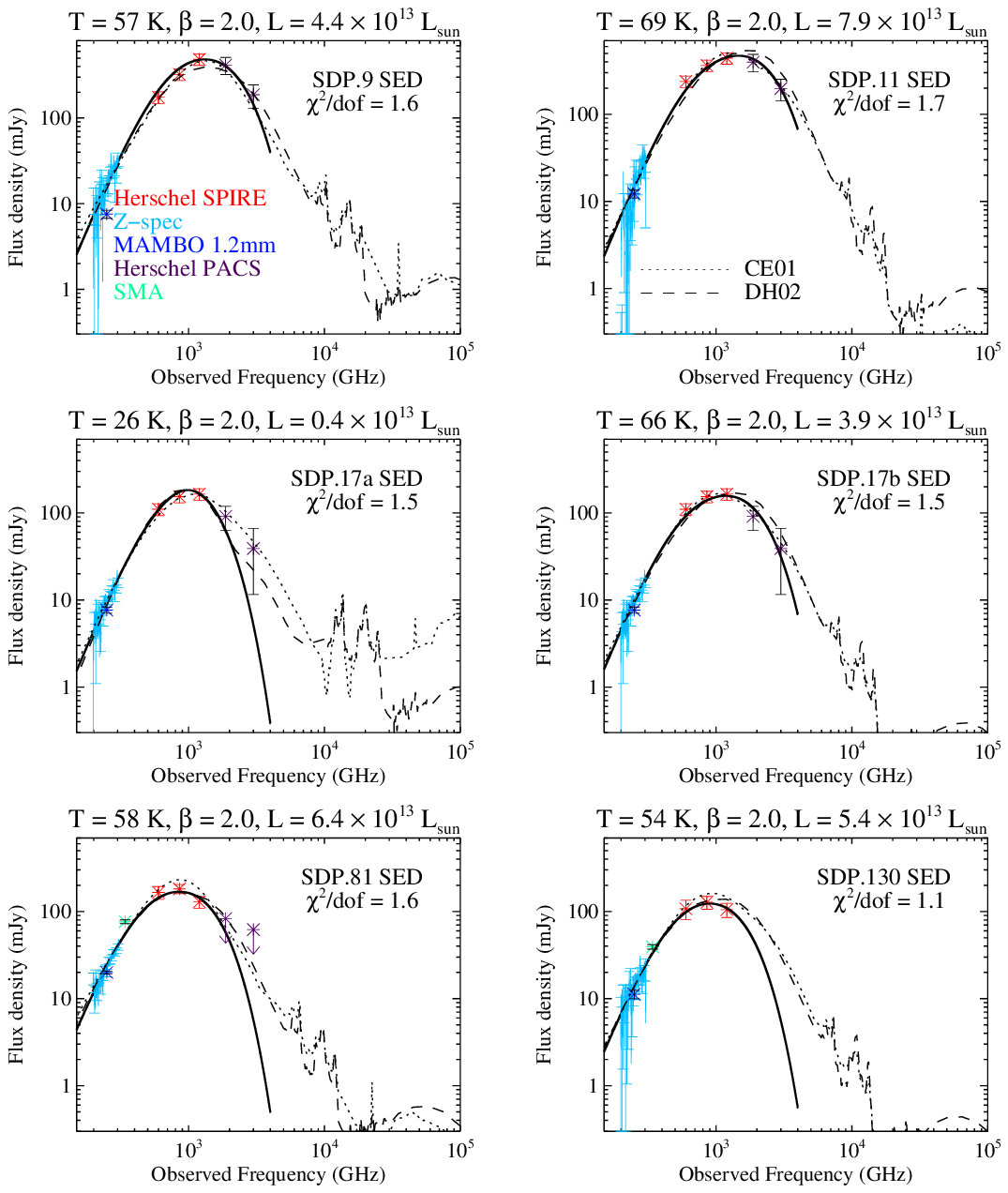}
\caption{The best-fit SED models for the five H-ATLAS galaxies in our sample. The continuous line shows the modified blackbody spectrum with $\nu_{0}=1300$~GHz and $\beta=2.0$, while the dotted and dashed lines show the SEDs obtained from the SED libraries of CE01 and DH02, respectively. The total infrared luminosities are calculated as the average between the CE01 and DH02 SED template fits, to account for emission above the blackbody spectrum at higher frequencies. The parameters for the modified blackbody fits are also listed in Table~\ref{tab:cont}.\label{seds}}
\end{figure*}

\subsection{Continuum Spectral Energy Distributions}\label{sseds}

The continuum data for all 5 galaxies is shown in Figure~\ref{seds}. The measured continuum flux from Z-Spec  is found to be in good agreement with the MAMBO 1.2~mm photometry \citep{Negrello10}, except for SDP.9. Estimates of the total amount of dust and star formation rates in each galaxy can be obtained by fitting their far-infrared (far-IR) to submillimeter spectral energy distribution (SED). In this fit we include the Z-Spec data along with the Herschel-SPIRE and Herschel-PACS photometric points, as well as the Submillimeter Array (SMA) measurements at 880~$\mu$m for SDP.81 and SDP.130 \citep{Negrello10}. 

The far-IR rest frame SED can be described by a modified blackbody function, defined as

\begin{equation}\label{bbody}
\begin{split}
F_{\nu}&=Q_{\nu}(\beta)B_{\nu}(T_{d})\Omega_{d}\\
&=(1-e^{-\tau{(\nu_{0})}(\nu/\nu_{0})^{\beta}})\frac{2h\nu^{3}}{c^{2}}\frac{1}{e^{h\nu/kT_{d}}-1}\Omega_{d}\\
&=\frac{L_{IR}}{4\pi d^{2}}\frac{Q_{\nu}(\beta)B_{\nu}(T_{d})}{\int Q_{\nu}(\beta)B_{\nu}(T_{d})d\nu},
\end{split}
\end{equation}

\noindent where $Q_{\nu}=1-e^{-\tau{(\nu_{0})}(\nu/\nu_{0})^{\beta}}$ is the emissivity, $B_{\nu}$($T_{d}$) is the Planck function, $\tau(\nu_{0})=1$ is the optical depth at $\nu_{0}$, $\Omega_{d} $ represents the observed solid angle of the dust emitting region, $d$ is the (known) distance to the source, and $h$ and $k$ denote the Planck and Boltzmann constants, respectively. The fit can be performed with three parameters: $T_{d}$, $\beta$, and a scale factor, while keeping $\nu_{0}$ constant. Including $\nu_{0}$ as a fourth parameter in the fit leads to a value of 1251$\pm$130~GHz for SDP.9, but no strong constraints are found for the rest of the sample, leading us to fix the $\nu_{0}$ at 1300~GHz. The low value found for $\nu_{0}$, and the observed flattening of the peak of the SEDs in the far-IR suggest that the SEDs of the galaxies in our sample can be modeled either as combinations of multiple graybodies with different temperatures, or as a single graybody with a large optical depth at far-IR wavelengths \citep{Papadopoulos2010b}.  The overall scale of the SED can be parametrized either in terms of the solid angle $\Omega_{d}$, or the total infrared luminosity ($L_{IR}$), defined as the integral of the SED from 8 to 1000~$\mu$m (rest frame). The $L_{IR}$ derived in this manner underestimates the true total infrared luminosity, due to the likely presence of warmer dust components that contribute at shorter wavelengths. 

The simplest model that can reproduce the data for the entire sample has fixed $\beta=2$ \citep{Priddey2001} and the already-mentioned $\nu_{0}$=1300~GHz, in agreement to the value found for SDP.9. The dust emissivity index $\beta=2$ is also consistent with the error bars of the Z-Spec spectra. The best-fit models are shown in Figure~\ref{seds}, and the corresponding values for $T_{d}$ are listed in Table~\ref{tab:cont}. With dust temperatures between 54 and 69~K, the peak of the restframe dust SED is found in a narrow range of wavelengths (73 to 92~$\mu$m) for all lensed galaxies in the sample. It is important to bear in mind that this fitting function for the SED is largely empirical, and the degree to which $T_{d}$ and $\beta$ represent physical quantities is complicated by the spatial averaging over the entire galaxy and the degeneracy between a distribution in dust temperature and a distribution of dust types (represented by $\beta$). The formal errors for the fitted parameters ($T_{d}$) should not be interpreted as errors on physical quantities, due to these caveats.

Even though such SED fits could be obtained using just the photometric points, the addition of Z-Spec data not only strongly constrains the continuum slope, but also breaks the degeneracy between $T_{d}$ and redshift \citep{Blain:1999}, by independently determining the latter. Since Z-Spec has determined the redshift, we are able to obtain $T_{d}$ from the continuum fit, which otherwise would constrain only the quantity $T_{d}/(1+z)$ \citep[e.g., ][]{Amblard10}. This degeneracy can lead to significant variations in the derived $T_{d}$ if the redshift is not measured independently. The implications of the continuum slope measured by Z-Spec for the dust composition is left for a future work.

Using Eq.~\ref{bbody} corrected for redshift and the derived $T_{d}$, we can estimate the observed size of the dust emitting region. This solid angle will be affected by the lensing magnification factor. If the dust optical depth at submm wavelengths is low, as is often the case, $\Omega_{d}$ will be correlated with $\tau$, and therefore with $\beta$ \citep{Hughes1993}. However, we break this degeneracy by fixing $\beta$. The resulting $\Omega_{d}$ ranges between 0.30~arcsec$^{2}$ for SDP.17b and 1.44~arcsec$^{2}$ for SDP.17a. These values may underestimate source sizes that are resolved in the SMA images with a resolution of $\sim$0.8~arcsec at 340~GHz \citep{Negrello10}. Using the magnification factors from Table~\ref{tab:cont}, the intrinsic size of the dust emitting region will have an equivalent radius of 0.7~kpc for SDP.81 and 1.3~kpc for SDP.130. Note however, that $\Omega_{d}$ corresponds to the effective solid angle of the dust emitting region, such as the total area of small clumps spread over a larger region. In an image where these clumps are unresolved, the total observed solid angle can appear to be larger. 

Having estimated the source size, the total dust mass follows from the relationship $\tau(\nu)=\kappa(\nu)M_{d}/D_{A}^{2}\Omega_{d}$, where $\kappa(\nu)$ is the dust absorption coefficient   $\kappa(\nu)=0.4(\nu/250GHz)^{\beta}$~cm$^2$~g$^{-1}$  \citep[e.g., ][]{Weiss2007}, and D$_{A}$ is the angular diameter distance. The dust mass can also be estimated in the optically thin limit ($1-e^{-\tau}\simeq\tau$) without the additional step of deriving $\Omega_{d}$, by substituting $\tau(\nu)$ directly in Equation~\ref{bbody}. This is a good approximation at 250~GHz (1.2~mm), in the middle of the Z-Spec bandpass. Calculated in the optically thin limit, the dust mass is a robust estimate of the lower limit for the total dust mass in the galaxy, $M_{d,lim}$. Using the optically thin approximation and the 250~GHz flux density measured by Z-Spec, we derive values for the magnified $M_{d,lim}$ of a few$\times$10$^{9}$~M$_{\odot}$, as listed in Table~\ref{tab:cont}. If the dust is optically thick, as suggested by $\nu_{0}$=1300~GHz, the calculated $M_{d,lim}$ will underestimate the true dust mass for our galaxy sample by at most 30\%. The dust mass is also inversely correlated with the assumed temperature, and will be underestimated when using the dust temperature corresponding to the peak of the SED. This temperature is likely too large to represent the bulk of the dust. Assuming that the 250~GHz flux is partially due to a dust component with a temperature as low as 20~K, and taking into account the optical depth corrections, we estimate that the total dust mass could be larger than $M_{d,lim}$ by up to a factor of $\sim$4. To summarize, with good approximation, the true dust masses for these galaxies will be found in the interval $[1,4]\times M_{d,lim}$. The remaining uncertainties in $M_{d,lim}$ are mostly due to uncertainties in the expression for $\kappa(\nu)$. Note that the quantity $M_{d}/\Omega_{d}$ is proportional to $\tau$ and independent of temperature; for a given $\tau$, a lower limit for $M_{d}$ implies a lower limit for $\Omega_{d}$, but this limit for $\Omega_{d}$ will decrease with increasing $\tau$.

A more realistic approach is to fit the photometric and continuum points with a library of SEDs, taking into account the transmission curve of each instrument. We apply this method to our galaxy sample, using the SED libraries of \citet{chary01} (CE01), and \citet{dale02} (DH02). The CE01 templates have also been used by \citet{Hwang2010} to fit both a low-z ($z<0.1$) and a high-z ($0.1<z<2.8$) galaxy sample, of which the highest luminosity tail seems to have properties overlapping the SMG population. Except for the subset of high-z galaxies with dust temperatures colder than $\sim$90\% of the local galaxies for a given luminosity, a subset that might be affected by blending, the CE01 template fits provide a good estimate for the total $L_{IR}$ in the high-z sample. For our lensed SMGs, we constrain well the peak of the SED and the dust temperature, and we are not in the regime where template mismatch can have a big impact on the inferred $L_{IR}$ \citep[see also ][]{Murphy2011}. 

We find that the IR luminosities derived from the modified blackbody fitting are at most a factor of $\sim$2 lower than those when we use the SED libraries, and within 20\% from the $L_{IR}$ obtained assuming the models of \citet{dacunha08}, calibrated for ULIRGs, with $A_{V}>2$ \citep{Negrello10}. The variations between the values of $L_{IR}$ obtained by different methods reflect the systematic uncertainties in deriving this quantity. Similar underestimates have been found by others, and are due to the fact that the submm photometry does not measure the warm dust component of the SED, if one is present \citep[e.g., ][]{Swinbank2010,Ivison10}. This is emphasized by the poor fit of the modified blackbody curve to the Herschel-PACS data points, and the significant flux at shorter wavelengths predicted by the CE01 and DH02 models (Figure~\ref{seds}). Any derivation of $L_{IR}$ is model dependent, with the largest differences arising from the presence of a warm dust component in the SED libraries. In Table~\ref{tab:cont} we list the $L_{IR}$ values derived from the SED template fitting method, as they represent a more accurate description of the total infrared energy output than the modified blackbody. The listed $L_{IR}$ are calculated as the average between the values given by the best-fit CE01 and DH02 templates. We find that the two best-fit templates give values for the $L_{IR}$ within 15\% of each other, both falling easily within our quoted error bars. On average, our $L_{IR}$ values are about 25\% lower than those found by \citet{Negrello10}, but these differences are difficult to judge without data shortward of 100~$\mu$m. Note that these values are rather smaller than typical IR luminosities of classical SMGs and unbiased sources detected by Herschel surveys.

\begin{deluxetable*}{cccccccccc}[h]
\tabletypesize{\footnotesize}
\tablewidth{0pt}
\tablecaption{\label{tab:sfr}Derived starburst properties and LTE parameters for the H-ATLAS galaxy sample.}

\tablehead{{H-ATLAS} & {$\mu$~$L_{CO}^{'}$} & {$\mu$~$L_{CO,corr}^{'}$} & {$\mu$~$M_{gas}$\tablenotemark{(a)}} & {$t_{SF}$\tablenotemark{(a)}} & {$N_{CO}$\tablenotemark{(b)}} & {$T_{ex}$\tablenotemark{(c)}} & {$\Omega_{s}/\Omega_{a}$} & {$\tau_{CO}$} & {$M_{CO}$\tablenotemark{(d)}}\\
{SDP ID} &(10$^{10}$ K km s$^{-1}$ pc$^{2}$) &(10$^{10}$ K km s$^{-1}$ pc$^{2}$) & (10$^{11}$ M$_{\odot}$) & (10$^{7}$ yr) & (10$^{17}$ cm$^{-2}$) & (K) & (10$^{-3}$) & &(10$^{6}$ M$_{\odot}$) }
\startdata
SDP.9   & 13$\pm$3  & 16$\pm$3 & 2.1 & 3.2 & 23$\pm$4 & 97$\pm$66 & 0.3 & 0.225 & 1.6$\pm$0.3 \\
SDP.11  & 15$\pm$5  & 18$\pm$6 & 2.4 & 2.1 & 47$\pm$15 & 48$\pm$8 & 0.3 & 0.929 & 3.2$\pm$1.0 \\
SDP.17a  & 4$\pm$3 & 5$\pm$3 & 0.7 & 11.1 & 3$\pm$1 & 160$\pm$90 & 1.3 & 0.005 & 6.5$\pm$2.2 \\
SDP.17b  & 12$\pm$3  & 16$\pm$4 & 2.1 & 3.7 & 22$\pm$6 & 80$\pm$17 & 0.3 & 0.114 & 1.5$\pm$0.4 \\
SDP.81 & 10$\pm$3  & 15$\pm$5 & 2.0 & 2.1 & 12$\pm$4 & 62$\pm$8 & 0.8 & 0.133 & 0.8$\pm$0.3 \\

\enddata
\tablecomments{The columns list: 1) the ID of the source in the SDP H-ATLAS catalogue; 2) the integrated brightness temperature of the lowest $J$ CO transition measured, times the source area; 3) same as Column 2, corrected for dust absorption (see Section~\ref{sco}); 4) the molecular gas mass; 5) the gas depletion time; 6) the CO column density; 7) the CO excitation temperature under LTE; 8) the estimated beam filling fraction for the lowest $J$ transition measured; 9) the optical depth for the lowest $J$ transition measured in that source. The parameters in the last four columns have been derived in the LTE approximation; and 10) estimated total mass of CO gas.}
\tablenotetext{(a)} {The errors for these parameters depend mostly on the uncertainties in the assumed conversion factors (see Section~\ref{sco}).}
\tablenotetext{(b)} {The values displayed correspond to an intrinsic source diameter of $\sim$2~kpc. The listed errors reflect the uncertainties in the measured integrated line fluxes. Other errors for this parameter, aside from the LTE model assumption, depend on our knowledge of the true source size.}
\tablenotetext{(c)} {The formal errors bars underestimate the uncertainty in $T_{ex}$, due to model assumptions restricted to LTE. In non-LTE models, a large region of the parameter space is allowed, and $T_{ex}$ becomes $J$-dependent (see Section~\ref{snlte}).}
\tablenotetext{(d)} {Corresponds to the assumed source radius of 1~kpc, except for SDP.17a which would have an estimated radius of 5.5~kpc at z=0.942, estimated from the optical image.}
\end{deluxetable*}

Except for SDP.17, we attribute all the submm flux density to the high redshift galaxy. The foreground lenses for the other galaxies in our sample have optical properties consistent with being quiescent elliptical galaxies, and are therefore unlikely to have a significant submm emission. We have attempted a decomposition of the SDP.17 SED, using the two measured redshifts and a wavelength-independent scaling factor for each of the two components. The $\chi^{2}$ value for the SED template fits is minimized when the observed flux density is split in half between the two components. This factor has been taken into account in Figure~\ref{seds} and in deriving the $L_{IR}$ for SDP.17a and SDP.17b, as listed in Table~\ref{tab:cont}. However, the large dust mass inferred for SDP.17a could be an indication that the SED decomposition between SDP.17a and SDP.17b overestimates the contribution of SDP.17a.  

We estimate the star formation rates for our galaxy sample using the conversion factor $SFR $($M_{\odot}$ yr$^{-1}$)=1.5$\times$10$^{-10}$($L_{IR}$/$L_{\odot}$) \citep{Solomon2005}, similar to the \citet{Kennicutt1998b} relation for a continuous starburst with a Salpeter initial mass function (IMF). Since the selected galaxies are lensed by foreground objects with magnification factors $\sim$10 \citep{Negrello10}, the intrinsic IR and CO line luminosities will be $\sim$10 times lower than the direct conversion from the measured fluxes. SDP.81 and SDP.130 have magnification factors of 25 and 6, respectively, as derived from the best-fit lens model to the high-resolution sub-mm images available for these two objects \citep{Negrello10}. In Tables~\ref{tab:cont} and \ref{tab:sfr}, we left the quantities affected by gravitational lensing magnification unmodified, for reference, but the presence of this contribution is indicated by the letter $\mu$ in front. Based on model predictions \citep{negrello07}, a typical amplification factor of 10 can be applied to these values. Once corrected for magnification, the infrared luminosities and corresponding SFRs are those typical of ULIRGs. 

\subsection{CO Line Luminosities and Spectral Energy Distributions}\label{sco}

The measurements of CO lines reveal important information about the physical properties and excitation conditions of the molecular gas, as well as the total gas budget in these galaxies. These parameters can be used to investigate the link between star formation and gas properties. Higher gas temperatures and lower densities would result in the increase of the Jeans mass, suggesting that star formation is biased towards high-mass stars \citep[e.g., ][]{Elmegreen2008,Klessen2007}. An increasing number of studies show that star formation may proceed differently in merger/starburst systems versus quiescent/disk systems, the former being characterized by a top-heavy initial mass function (IMF) \citep{Weidner2011,Habergham2010}. Such an IMF not only arises in dense starburst environments, but also has been invoked to explain the observed number counts at 850~$\mu$m \citep{Baugh2005}. A top-heavy IMF can arise in high-density material, shielded from far-UV radiation, but permeated by cosmic rays or X-rays, which heat the gas efficiently and generate cosmic-ray dominated regions \citep[CDRs, ][]{Papadopoulos2010imf} or X-ray dominated regions \citep[XDRs, ][]{bradford09,Schleicher2010}. The presence of a top-heavy IMF would have important consequences for the SFR inferred from the total $L_{IR}$. However, the XDR signatures, such as highly-excited CO lines \citep[e.g. ][]{bradford09}, will likely indicate that the galaxy is dominated by the presence of an AGN, but not directly probe the IMF. Since the gas properties derived from the analysis of the CO lines are galaxy-averaged, only with multiple CO lines we can begin to disentangle different PDR and XDR contributions \citep[e.g. ][]{vanderWerf2010}, which may help us characterize the star formation in these galaxies. Further understanding would require spatially resolving the star-forming regions, and probing the high-density star-forming gas with additional molecular tracers.

In order to derive the physical characteristics of the gas in these galaxies, including the gas temperature, density, pressure, and CO column density, we need measurements of multiple CO transitions, sampling the rotational ladder as fully as possible. The spectral line energy distribution (SLED) for the CO molecule has been constructed in a few cases for nearby and low redshift galaxies \citep[e.g., ][]{Panuzzo2010}. In Figure~\ref{sleds} we show the partial SLEDs for our galaxy sample, constructed from the lines detected in the Z-Spec bandpass. This plot favors a distribution with the brightest lines between CO(5$-$4) and (7$-$6), similar to the distribution observed for other SMGs and starburst galaxies \citep{Weiss07,Danielson2010}. 

The shape of the line luminosity distribution does not reflect only the gas kinetic temperature, but also the gas density, and the effects of the optical depth at the line frequency \citep{Goldsmith1999,Papadopoulos2010b}. In the optically thin limit, the CO column density scales with the absolute value of the line intensity, assuming that the source size and the magnification factor are known. Under the assumption of local thermodynamic equilibrium (LTE), all CO transitions have the same excitation temperature, $T_{ex}$, also equal to the gas kinetic temperature $T_{kin}$, signifying that all rotational levels are populated according to the Maxwell-Boltzmann distribution at temperature $T_{ex}$. In Section~\ref{slte} we estimate these parameters by fitting the partial SLEDs, using the relationship between the integrated line brightness temperature, column density and excitation temperature, under LTE. Although this case is limiting due to the assumption of constant $T_{ex}$ for all levels, it is interesting to compare the predictions of this model to the more general non-LTE models, given its simple physical interpretation. In the non-LTE case, presented in Section~\ref{snlte}, the models involve a larger number of parameters, and are less well constrained. We use RADEX \citep{vanderTak07} to compute the brightness temperatures of the CO lines and estimate the likelihood distribution over the parameter space. These distributions allow us to assess if the available data are able to distinguish between the LTE and non-LTE models.

\subsubsection{Gas Masses}\label{gmass}

A useful quantity describing the CO lines is the velocity-integrated brightness temperature scaled by the area of the source, $L_{CO}^{'}$, in units of K~km~s$^{-1}$~pc$^{2}$. If the CO is thermalized and the lines are optically thick, $L_{CO}^{'}$ will be the same for all rotational transitions for which the Rayleigh-Jeans approximation holds. In what follows, the brightness temperatures are computed in the Rayleigh-Jeans limit, and the values for $L_{CO}^{'}$ are listed in Table~\ref{tab:sfr} both corrected and un-corrected for dust absorption. Taking into account our estimate of the dust optical depth in Section~\ref{sseds}, the observed brightness temperature of the CO lines will be related to the intrinsic brightness temperature via the relation $T^{obs}=\mathrm{exp}(-\tau_{d})T^{int}$, where $\tau_{d}=(\nu/\nu_{0})^{\beta}$. This correction will tend to boost the intensities of the higher $J$ lines and drive the excitation of the gas higher \citep[see also][]{Papadopoulos2010b}. The physical parameters of the gas derived in Sections~\ref{slte} and \ref{snlte} are also based on the absorption-corrected line intensities, and the effects of this correction are discussed as necessary. 

$L_{CO}^{'}$ is traditionally derived from the CO(1$-$0) transition and related to the total molecular mass via the empirical relation $M_{gas}=\alpha L_{CO}^{'}$, where $\alpha$ is 4.6~M$_{\odot}$~(K~km~s$^{-1}$~pc$^{2}$)$^{-1}$ for the Galaxy \citep{Solomon1997}, and 0.8~M$_{\odot}$~(K~km~s$^{-1}$~pc$^{2}$)$^{-1}$ for ULIRGs \citep{Downes1998,tacconi08}. Following \citet{Solomon2005}, we use the latter value for $\alpha$ and the $L_{CO}^{'}$ for the lowest observed CO transition to determine the gas masses (see Table~\ref{tab:sfr}). This procedure assumes that all transitions from CO(1$-$0) up to the lowest observed are thermalized, which might not necessarily be the case. A recent comparison of the CO(3$-$2) and (1$-$0) lines \citep{Harris10} shows that the ratio of the brightness temperatures for these two lines averages to 0.6 rather than 1, due to the presence of multi-phase CO gas. Moreover, the mid-$J$ CO transitions do not account for the possible presence of a colder gas component, making the $M_{gas}$ derived in this manner a lower limit for the total gas mass in the galaxy.  Assuming that the lines with  $J_{u}>$3 are thermalized, corresponding to a warmer gas component, we apply this correction factor to the lowest CO transition measured, and obtain the gas masses listed in Table~\ref{tab:sfr}. However, for subthermal excitation the ratio between the brightness temperatures of higher-$J$ CO lines and CO(1$-$0) could be even smaller. The value of $L_{CO(1-0)}^{'}$ for SDP.81 derived from the CO(1$-$0) line intensity is $1.8\times 10^{10}$~K~km~s$^{-1}$~pc$^{2}$, after correcting for the lensing magnification factor \citep{Frayer10}, which results in a brightness temperature ratio between the CO(7$-$6) and CO(1$-$0) lines of 0.33$\pm$0.16. This also indicates that our conversion factors will globally underestimate the total gas mass.

Using the SFRs derived from the IR luminosities, the gas reservoir probed by CO implies a gas depletion time [$M_{gas}/SFR$] in these objects of $\sim10^{7}$ years, similar to other known SMGs \citep{Solomon2005,greve05}. This can be interpreted as the starburst lifetime under the assumptions of constant SFR and no gas inflow. Note that in the absence of differential lensing, this estimate of the gas depletion time is independent of lensing magnification. We currently do not have enough data available to construct lensing models and constrain the differential lensing for each of these sources. The star formation efficiency can be expressed directly in terms of $L_{IR}/L_{CO}^{'}$, without the need for a gas mass or SFR conversion factor. After accounting for the lensing magnification factor, $L_{IR}$ and $L_{CO}^{'}$ for our sample follow the same relationship as other SMGs and ULIRGs \citep{greve05,Wang2010}, within the scatter. 
 
We derive an average molecular gas-to-dust ratio for the lensed galaxies of 127$\pm$50, subject to the caveats above: the gas mass is underestimated using the standard conversion factor, and the dust mass is also underestimated by the single component model fit. The mean gas-to-dust ratio does not include the foreground SDP.17a, and is in agreement with the values found for other SMG samples \citep{Kovacs2006,Michalow2010,Santini2010}. Similar to the gas depletion lifetime, this ratio will be independent of magnification if we ignore differential lensing.

\subsubsection{LTE models}\label{slte}

The integrated line flux $S_{\nu}\Delta$v (in Jy~km~s$^{-1}$) in the observer's frame is related to the velocity-integrated Rayleigh-Jeans source brightness $W(J)$ by \citep[e.g., ][]{Solomon1997}

\begin{equation}\label{ttojy}
W(J)=\frac{\lambda^{2}_{J,J-1,rest}(1+z)^{3}}{2k \Omega_{a}}S_{\nu}\Delta\mathrm{v}\frac{\Omega_{a}}{\Omega_{s}}
\end{equation}

\begin {figure}
\centering
\includegraphics*[scale=0.35,angle=90]{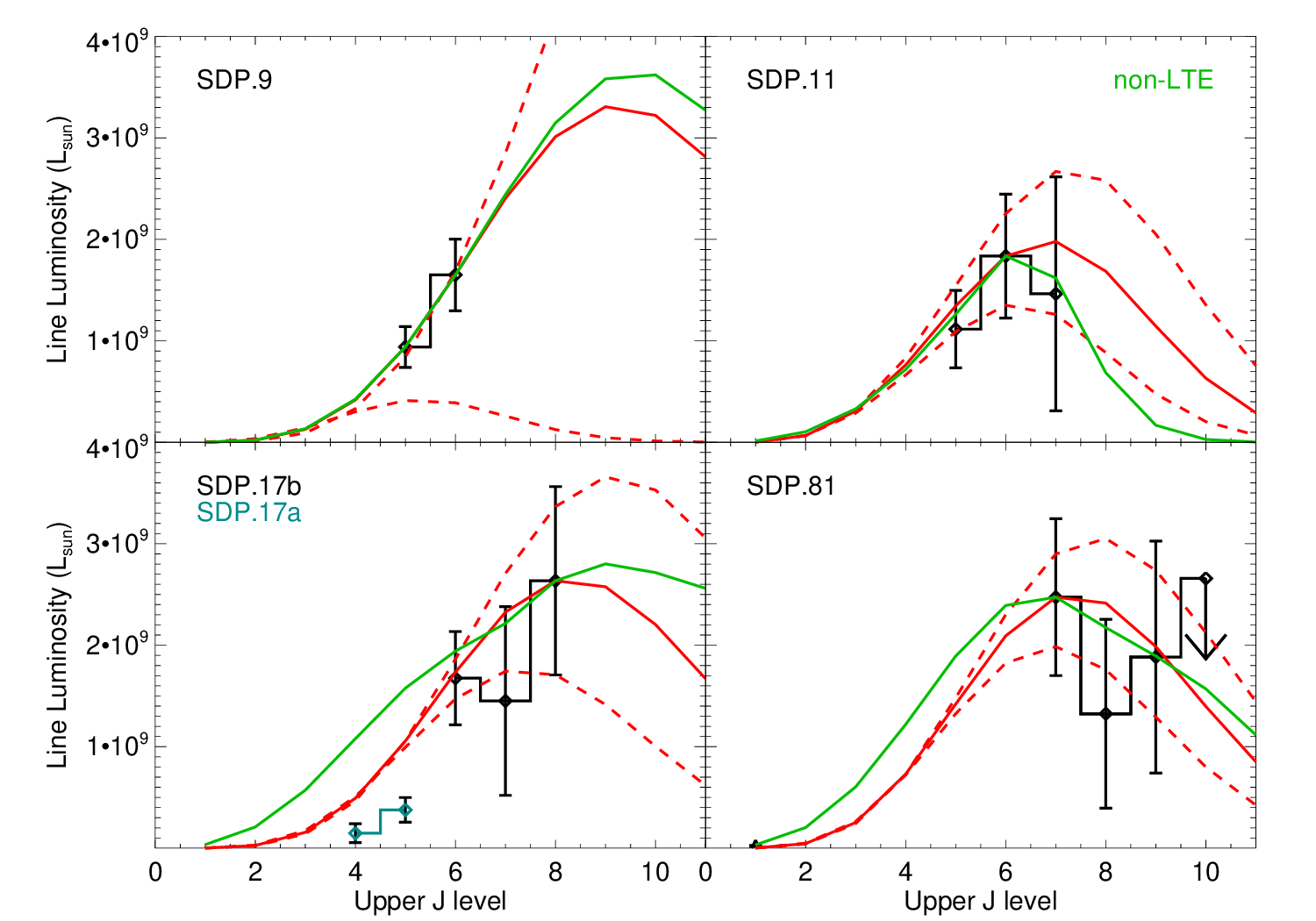}
\caption{Spectral line energy distributions, uncorrected for gravitational lensing magnification. The Z-Spec measurements are shown connected by the black histogram. The data point for the CO(1$-$0) line measured by \citet{Frayer10} in SDP.81 falls at the bottom of the panel, and is better seen in Figure~\ref{tempsone}. The red lines show the SLEDs predicted by the best-fit LTE model (continuous), and the LTE models corresponding to the limits of the 1$\sigma$ standard confidence interval for $T_{ex}$ determined from the fit (dashed). The green line shows the SLED predicted by RADEX, using the parameters corresponding to the 4D maximum likelihood solution, as listed in Table~\ref{tab:radex}.  \label{sleds}}
\end{figure}

\noindent where $\Omega_{s}$ and $\Omega_{a}$ are the solid angles of the source and the antenna, respectively. $W(J)$ is in units of K~km~s$^{-1}$. The last fraction represents the inverse of the beam filling fraction. The contribution of the gravitational lensing magnification should cancel out in this expression, as it contributes to both $S_{\nu}$ and $\Omega_{s}$, but the true $\Omega_{s}$ is not known. In principle, the same approach taken for the continuum (Section~\ref{sseds}) could be used to determine the source size. However, such a fit requires a minimum of three parameters, and will not be well constrained by the number of CO lines in our SLEDs. In addition, the optical depth depends directly on the column density, and cannot be estimated independently, in the same way that the dust optical depth was determined by the continuum slope. We assume an intrinsic source size of $\sim$2~kpc, consistent with the angular diameter of 0.2\arcsec\ of the SMG SMMJ2135-0102 at z=2.3259 \citep{Swinbank2010,Danielson2010} used by \citet{Negrello10} for the SDP H-ATLAS sources, and similar to the size of the dust emitting region found in Section~\ref{sseds}. The corresponding beam filling fractions are listed in Table~\ref{tab:sfr}. As this source solid angle now represents the intrinsic size, and not the magnified one, we must correct the observed flux densities by the lensing magnification factors. We use the values listed in Table~\ref{tab:cont}, when available, and assume a value of 10 in all other cases. The case of SDP.17a is treated differently, as it is assumed to be a foreground galaxy, not affected by gravitational lensing. For the intrinsic size of SDP.17a, we use a value of 1.54~arcsec$^{2}$, which approximates the size of the optical image. \citet{Negrello10} identify two galaxies in the i-band image of SDP.17 and fit both light distributions with the GALFIT software. As the presence of two galaxies could indicate a possible merger, we choose the source size of SDP.17a to be the sum of the areas of these two galaxies.

The distribution of the velocity-integrated brightness temperatures for the CO lines can be constructed starting from the CO column density and gas temperature, under the assumption of LTE. Following \citet{Goldsmith1999}, the velocity-integrated Rayleigh-Jeans source brightness is given by

\begin{equation}\label{btemp}
W(J)=N_{J}\frac{hc^{3}A_{J,J-1}}{8\pi k\nu^{2}}\frac{1-e^{-\tau_{J,J-1}}}{\tau_{J,J-1}},
\end{equation}

\noindent where $\tau_{J,J-1}$ is the line center optical depth, and $A_{J,J-1}$ is the Einstein $A$ coefficient for the transition. In LTE, the column density of molecules in the upper level, $N_{J}$, is related to the total column density $N$, by
\begin{equation}\label{nupper}
N_{J}=\frac{N}{Z}g_{J}e^{-E_{J}/kT_{ex}},
\end{equation}

\noindent where $Z$ is the partition function, $E_{J}$ is the energy of level $J$, and $g_{J}=2J+1$ is the degeneracy of level $J$. 
The line center optical depth can be expressed as a function of column density, temperature, and line width $\Delta$v as

\begin{equation}\label{tauline}
\tau_{J,J-1}=A_{J,J-1}\frac{c^{3}}{8\pi\nu^{3}\Delta\mathrm{v}}N_{J}(e^{h\nu/kT_{ex}}-1).
\end{equation}

\noindent We fit Eq.~\ref{btemp} to the measured $W(J)$ distribution, with the column density and gas temperature as free parameters, and $\Delta$v~=~300~km~s$^{-1}$. We find that the best fit models have relatively low optical depths ($\lesssim 1$) such that the choice of the line width has only a small effect on the fitted parameters. For the lensed galaxies, the measured CO SLEDs and the range of SLEDs allowed by the formal 1$\sigma$ interval for the gas temperature are are shown in Figure~\ref{sleds}. Since the CO lines are found to be close to optically thin in this model, the column density 1$\sigma$ interval would only scale these curves up and down, and not affect their overall shape.

The SLEDs can be characterized by an overall scale and line ratios. The scale of the observed SLEDs is mainly a result of the CO column density and the beam filling fraction, while the line ratios depend on the CO temperature and gas (H$_{2}$) density. The parameters in each pair are therefore largely degenerate and anti-correlated. This degeneracy is characteristic to CO and other molecular SLEDs, regardless of galaxy type. The last correlation (between temperature and gas density) only exists until LTE is reached, and the temperature becomes fixed. By making assumptions on the beam filling fraction and gas density, we can place limits on the remaining parameters. The error bars on the column densities derived in this manner are correlated with the errors in the beam filling fraction, which are not known. Similarly, by making the assumption of LTE for all transitions up to CO~(7-6), we are constraining the gas density to be greater than the critical density for this transition ($n$[H$_{2}$]~$\gtrsim$~3$\times$10$^{5}$~cm$^{-3}$). At densities $n$[H$_{2}$]~$\gtrsim10^{6}$~cm$^{-3}$, considerably larger than the average value observed in Galactic molecular clouds, all observed lines should be in LTE. Values of the gas density more typical for Galactic molecular clouds (10$^{3}$-10$^{4}$~cm$^{-3}$) correlate with higher gas temperatures, of a few hundred degrees, in order to reproduce the observed line ratios.

The best-fit LTE CO column densities are $\sim$few$\times$10$^{18}$~cm$^{-2}$, and the gas temperature ranges between 48 and 160~K, as listed in Table~\ref{tab:sfr}, with the largest errors corresponding to the cases where only two CO lines have been measured. Note that these temperatures are derived after correcting the line fluxes for dust extinction, and are on average larger than the temperatures that would be obtained without correcting the line fluxes (between 41 and 115~K). However, due to the large errors in our measurements, these differences are not significant.

Taking into account the assumed source size, we estimate total CO masses (M$_{CO}$) of a few$\times$10$^{6}$~M$_{\odot}$, or $\sim$10$^{-4}$ of the total gas mass. This is consistent with the average relative abundance of CO, and would account for the entire molecular mass. However, since these LTE models imply very large pressures ($\sim$10$8$~K~cm$^{-3}$), the total molecular content of the galaxy would have to reside in dense star-forming cores, which would account for the total CO emission. This suggests that the LTE parameters cannot describe the overall average conditions of the gas in the galaxy. Other regions of the parameter space are associated with non-LTE gas excitation, explored with the RADEX modeling in the next section.

\subsubsection{Non-LTE radiative transfer models of CO line excitation}\label{snlte}

In general, the rotational levels of the CO molecule might not be populated according to a single temperature, and the CO excitation temperature does not equal the gas kinetic temperature. By dropping the LTE assumption, we allow the excitation temperature to be a function of transition, being determined by the level populations for each line, while the kinetic temperature will be the global quantity describing the thermal energy of the gas. The level populations are found by solving the detailed balance equations including both radiative and collisional rates, and the output intensities are calculated by solving the radiative transfer equations. Usually, these equations are strongly coupled, involving large spatial and frequency grids, and further complicated by the number of molecules and transitions involved. Simplifying assumptions are usually made to reduce the computing time, depending on the problem at hand. 

We use RADEX to estimate the range of physical parameters consistent with the measured line strengths when dropping the LTE assumption. RADEX is a one dimensional, non-LTE radiative transfer code, that solves for the level populations iteratively, employing the escape probability approximation for the radiative transfer \citep{vanderTak07}. The medium is assumed homogeneous and isothermal, and the number, type, and abundance of the participating molecules is selectable by the user. The input parameters are the kinetic temperature, $T_{kin}$, the number density of molecular hydrogen, $n$[H$_{2}$], as the collisional partner, and the column densities per unit line width of the participating molecules, only CO in our case. The background radiation field is the cosmic microwave background (CMB), redshifted according to the redshift of each galaxy. The output contains the predicted line excitation temperatures, optical depths, and line intensities. The output line fluxes are scaled by an additional factor $\phi$, that represents fractional corrections to the size of the emitting region and to the gravitational lensing magnification factor. It would correspond to the area filling fraction of the emitting region, if the size and lensing magnification factor of the source were known precisely. A value $\phi>1$ would suggest that the assumed source size was underestimated. We compare the measured flux densities with the line intensities output by RADEX using the same values for source sizes, line widths, and lensing magnification factors assumed in Section~\ref{slte} for the LTE model. For the case $\phi=1$ and $n$[H$_{2}$]$\gg n_{crit}$ for all transitions, RADEX will recover the LTE SLED as determined from $T_{ex}$ and N[CO] in Section~\ref{slte}, as expected. 

\begin{deluxetable*}{cccccccc}[h]
\tabletypesize{\footnotesize}
\tablecaption{\label{tab:radex}Parameters used for the RADEX models shown in Figures~\ref{sleds} and \ref{tempsone}.}
\tablehead{{RADEX} & {$T_{kin}$} & {log($N_{CO}$)} & {log($n$[H$_{2}$])} & {$\phi$} & {log($P$)}& {log($M_{gas}$)} & $L_{CO,total}^{'}$ \tablenotemark{(d)}\\
{Model} &(K) & (10$^{18}$ cm$^{-2}$) & (cm$^{-3}$) &  & (K cm$^{-3}$) & (M$_{\odot}$) & (10$^{10}$ K km s$^{-1}$ pc$^{2}$)}

\startdata
SDP.9\tablenotemark{(a)}   & 99.   & 18.79 & 7.63 & 0.44 & 9.6 & 8.96 & 12.3\\
68\% credible region\tablenotemark{(b)}   & 90.-1083.   & 18.78-20.96 & 5.39-7.97 & 0.03-0.63 & 7.44-10.39 & 8.81-9.88 & 5.5-37.3\\
\hline
SDP.11  & 24.  & 19.60 & 7.35 & 1.7 & 8.7 & 10.37 & 15.5 \\ 
68\% credible region   & 24-612   & 18.62-20.40 & 4.20-7.93 & 0.11-1.58 & 6.63-9.93 & 9.19-10.37 & 0.9-32.0\\
\hline
SDP.17b  & 2833.  & 19.73 & 2.31 & 0.91 & 5.76 & 10.24 & 24.4 \\
68\% credible region   & 154-2884  & 18.44-20.22 & 2.31-6.38 & 0.07-0.95 & 5.76-8.69 & 8.97-10.24 & 1.9-38.6 \\
\hline
SDP.81 & 375.  & 20.09 & 2.88 & 0.26 & 5.46 & 10.04 & 10.7 \\
68\% credible region   & 89-1416   & 18.92-20.91 & 2.88-6.13 & 0.01-0.29 & 5.45-8.28 & 8.57-10.04 & 0.2-31.2\\
SDP.81*\tablenotemark{(c)} & 453.  & 20.07 & 2.98 & 0.18 & 5.63 & 9.85 & 9.0 \\
68\% credible region   & 40-477   & 19.66-20.35 & 2.90-4.43 & 0.17-0.74 & 5.37-6.00 & 9.70-10.18 & 3.4-18.2\\
\enddata
\tablecomments{The columns list: 1) the model notation; 2) the kinetic temperature $T_{kin}$ (under LTE, $T_{kin}$=$T_{ex}$); 3) the CO column density; 4) the density of H$_{2}$; 5) $\phi$ is an overall scaling factor, that would correspond to the area filling fraction if the intrinsic source size and gravitational lensing magnification factor were known exactly. This enters as the fourth unknown parameter in the maximum likelihood estimation; 6) the gas pressure; 7) the total gas mass in the beam; and 8) the $L_{IR}$/$L_{CO,total}^{'}$ as a measure of the star formation efficiency predicted by each model, where $L_{CO}^{'}$ is summed over all CO transitions in the model.}
\tablenotetext{(a)}{These parameters correspond to the 4D maximum likelihood solution from an MCMC exploration of the parameter space with 10$^{5}$ iterations for each galaxy. Additional measured CO transitions would help rule out solutions with extreme temperatures and densities.}
\tablenotetext{(b)}{This represents the smallest interval enclosing 68\% of the $marginal$ probability for each parameter.}
\tablenotetext{(c)}{This second model for SDP.81 includes the CO(1$-$0) measurement from \citet{Frayer10}.}
\tablenotetext{(d)}{When derived from the integrated brightness temperature, a source radius of 1~kpc is assumed.}
\end{deluxetable*}

We run RADEX for a range of input models, parametrized by $T_{kin}$, $N$[CO]/dv, $n$[$H_{2}$] and $\phi$, and compute the likelihood density function for all models following the method described in \citet{Ward03}. Weak priors are set to rule out unphysical solutions, keeping the total molecular mass smaller than the dynamical mass, and the length of the CO column smaller than the physical size of the galaxy \citep{Ward03,Panuzzo2010}. The dynamical mass cut-off is estimated choosing the line width of 300~km~s$^{-1}$, and we require that the gas be self-gravitating \citep[$K_{vir}\geq 1$, e.g., ][]{scott10,Papadopoulos2004}. We also impose a limit for the kinetic temperature at 3000~K, where collisional dissociation begins to rapidly destroy CO, weakly dependent on the gas density. 

\begin {figure}[h]
\centering
\includegraphics*[scale=0.5,angle=0]{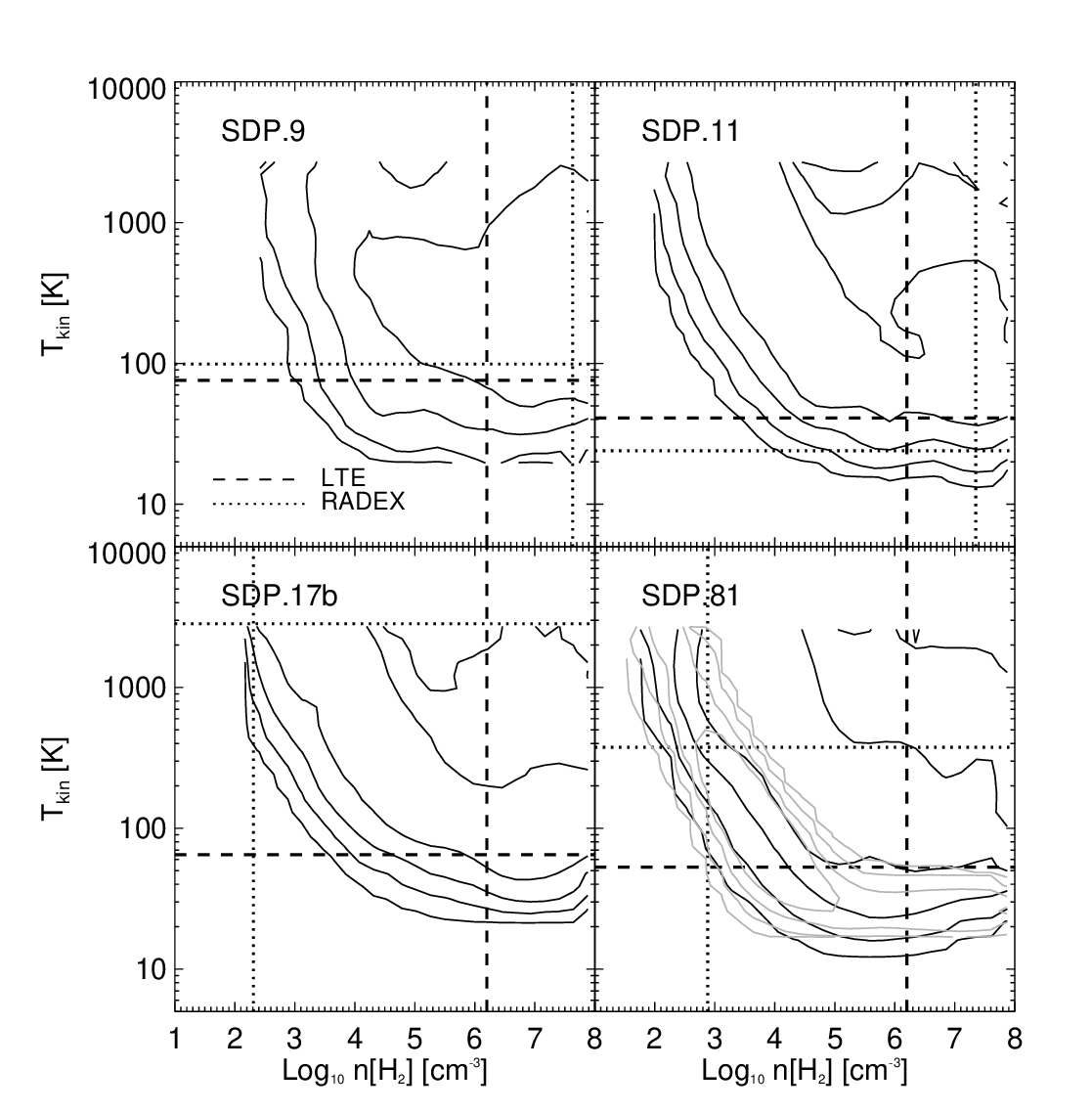}
\caption{Contour plots of the ($T_{kin}$, $n$[H$_{2}$]) 2D marginal likelihood distributions, generated by a MCMC sampling of the parameter space for RADEX models. The contours are in n$\sigma$-equivalent steps, enclosing 68.3\%, 95.4\%, 99.7\%, and 99.99\% of the probability, respectively. The dashed lines correspond to parameters that reproduce the LTE solution, and the dotted lines indicate the parameters corresponding to the RADEX 4D maximum likelihood solution. Note that the 2D marginal distributions will not necessarily have the same maximum as the 4D distribution. The kinetic temperature is limited to $< 3000$~K, where collisional dissociation of CO becomes important. In the SDP.81 panel, the lighter contours show the probability levels for a model including the CO(1$-$0) from \citet{Frayer10}. The parameters for this model are listed as model SDP.81* in Table~\ref{tab:radex}. \label{tnmcmc}}
\end{figure}

\begin {figure}[h]
\centering
\includegraphics*[scale=0.5,angle=0]{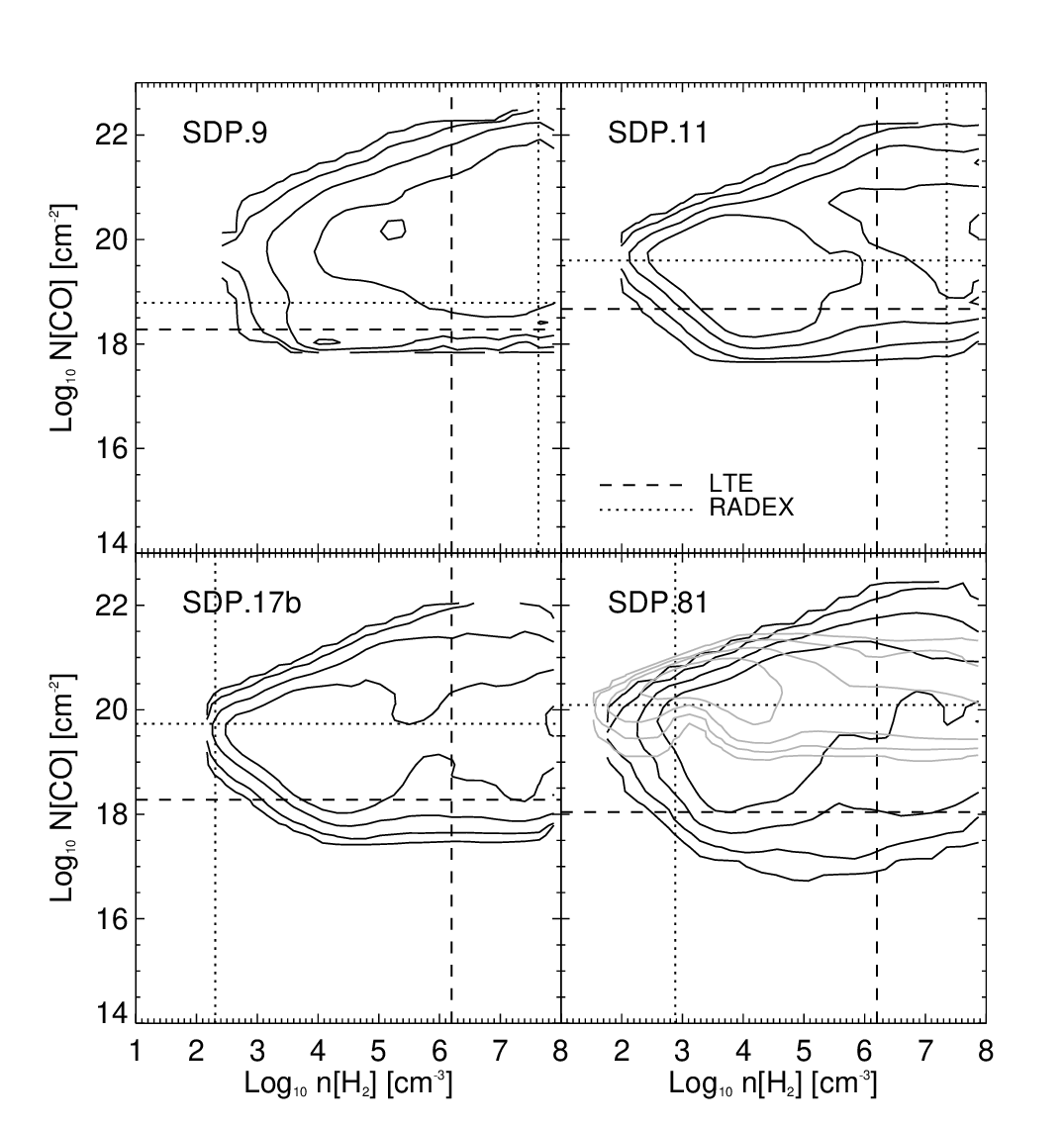}
\caption{Same as Figure~\ref{tnmcmc} for the ($N[CO]$, $n$[H$_{2}$]) 2D marginal likelihood distributions. \label{nnmcmc}}
\end{figure}

We map the surface of the likelihood distribution and determine the location of its maximum by running a Markov chain Monte Carlo (MCMC) algorithm, described in detail in \citet{scott10}. Due to the large error bars and small number of data points, the aforementioned priors have only a weak effect on the final result, and mostly prevent the MCMC from spending time exploring unphysical regions of the parameter space. The 2D marginal probability contours obtained from the MCMC algorithm are shown in Figures~\ref{tnmcmc} and \ref{nnmcmc}, with the position of the 4D maximum likelihood indicated by the dotted line. Note that the 4D probability distributions are highly non-gaussian, and therefore the coordinates of the maxima for the marginalized distributions in 2D do not match, in general, the paramters corresponding to the maximum of the 4D distribution. The set of parameters that maximizes the 4D likelihood for each galaxy is listed in Table~\ref{tab:radex}, and the line luminosities predicted by this model are shown in blue in Figure~\ref{sleds}. The 68\% credible regions are calculated as the smallest intervals containing 68\% of the 1D marginal probability for each parameter, around the value corresponding to the 4D maximum likelihood. In some cases, the credible regions for $\phi$ suggest that the size of the emitting region could be larger than assumed for the LTE models, interpreted as a larger area characterized by lower gas density and pressure than the LTE case.

\begin {figure}
\centering
\includegraphics*[scale=0.35,angle=90]{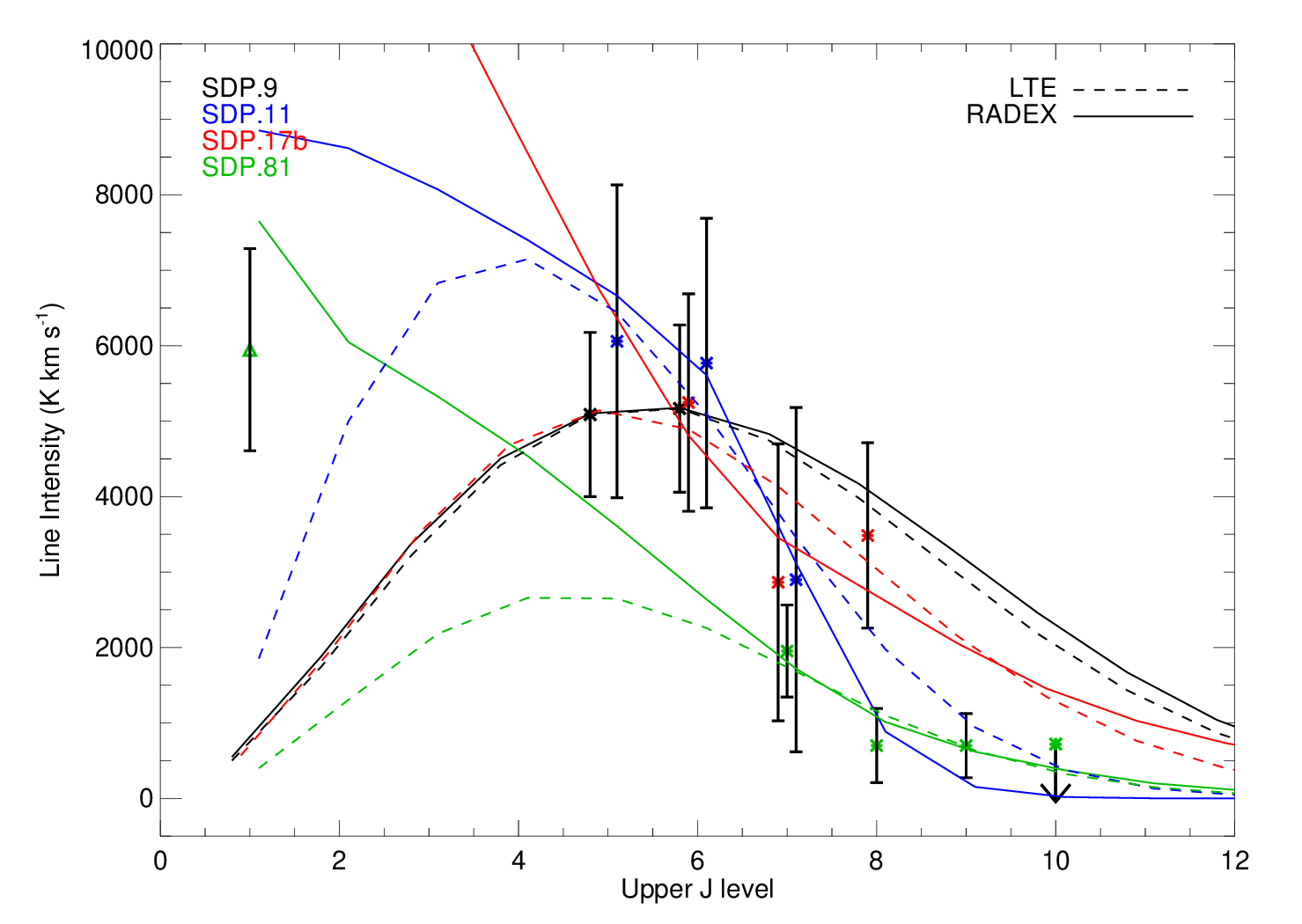}
\caption{$W(J)$ as a function of transition for four of the galaxies in our sample. For clarity, the points corresponding to the same transition in different galaxies have been slightly offset left and right from the position of the exact upper $J$ level. The triangle point represents the intensity of the CO(1$-$0) line for SDP.81 measured by \citet{Frayer10}. The $W(J)$ distributions predicted by the LTE and non-LTE models are shown with a dashed and continuous line, respectively. These lines emphasize the constraints on the allowed parameter space that can be gained by having measurements of both higher and lower $J$ transitions. While the Z-Spec data cannot clearly favor one of the models, the non-LTE model is superior when including the CO(1$-$0) line for SDP.81. \label{tempsone}}
\end{figure}

Due to the aforementioned degeneracies (see Section~\ref{slte}), the product of the kinetic temperature and gas density on one hand, and CO column density and $\phi$ on the other hand, are better constrained than individual parameters. These products are linearly proportional to the gas pressure and total gas mass, respectively, quantities which are listed in Table~\ref{tab:radex}. 

The gas mass has also been derived in Section~\ref{gmass}, using two parameters: 1) the conversion factor $\alpha=0.8$ between the gas mass and $L_{CO(1-0)}^{'}$ derived by \citet{Downes1998} from a non-LTE model, and 2) the the scaling between the brightness temperatures of higher $J$ CO lines and that of CO(1$-$0), $R_{J,1}=T_{B}^{J,J-1}/T_{B}^{1,0}$, using the value $R_{3,1}=0.6$ \citep{Harris10}, which we assumed to hold for higher $J$'s. For comparison, we can independently estimate both factors, $\alpha$ and $R_{J,1}$, using our best-fit non-LTE models. For $\alpha$ we get an average value of $\alpha=0.46\pm0.24$ for the whole sample, not taking into account the error bars in the best-fit parameters. While for $R_{3,1}$ we obtain an average value of $R_{3,1}= 0.73$, in relative agreement with the more reliable value of $0.64\pm0.1$ obtained by \citet{Harris10}, for the higher $J$ ratios we have $R_{5,1}= 0.75$ (SDP.11), $R_{6,1}= 0.26$ (SDP.17b), and for SDP.81 $R_{7,1}= 0.22$ and $R_{7,1}= 0.27$ for the 2 non-LTE models, respectively (see Table~\ref{tab:radex}). This suggests that for higher $J$ lines the $R_{J,1}$ factor may be closer to a value of 0.3 for these excitation conditions. By comparison to the models of \citet{Narayanan2009}, values of 0.3 are marginally allowed, on the high end of the range. We emphasize that these estimates are strongly model-dependent, and only direct measurements of the CO(1$-$0) lines would make it possible to both constrain the models and validate these values. It is becoming apparent that since most of our knowledge about local dust-enshrouded galaxies comes from the study of low-$J$ CO lines, while at high redshift the high-$J$ CO lines are more readily accessible, we need to be able to measure both in order to make a direct comparison of the excitation conditions and gas properties. Important progress in this direction, by measuring the high-$J$ lines in local galaxies, has been made with $Herschel$ in recent years \citep[e.g., ][]{Panuzzo2010, Rangwala2011}.

The region of the parameter space that is most consistent with the observed line strengths is enclosed by the likelihood contours in Figures~\ref{tnmcmc} and \ref{nnmcmc}. The likelihood space roughly splits into high density/low temperature, and low density/high temperature solutions. One additional complication to the interpretation arises from the high dust optical depths, which lead to the suppression of CO lines with increasing frequency, and will cause an underestimate of the excitation temperature if unaccounted for \citep[e.g., ][]{Papadopoulos2010b}. As mentioned in Section~\ref{slte}, we attempt to account for this effect by  correcting the CO line strengths for dust absorption using the dust optical depths estimated in Section~\ref{sseds}. However, the likelihood distribution is relatively shallow over the whole region, reflecting the insufficient amount of information in our data, and the likelihood contours are only marginally affected by this correction. 

To emphasize the insight gained by including additional lines in the fit, we add to the SLED of SDP.81 the CO(1$-$0) integrated flux from \citet{Frayer10}. A likelihood analysis for the new set of lines results in the best fit parameters listed in Table~\ref{tab:radex} as model SDP.81*. The 2D marginalized likelihoods for this case are shown by the light grey contours in Figures~\ref{tnmcmc} and \ref{nnmcmc}. The tightening of the likelihood contours is substantial with just one line added to the data, and the LTE region of the parameter space becomes less favored. However, the limitation of this model is that it assumes a single gas component, while most of the emission in the CO(1$-$0) line could be originating from cold molecular gas.

The constraints on the parameter space for the non-LTE models are weak, as expected given the limited sampling of the SLED and the large error bars, and cannot well distinguish between the LTE and non-LTE scenarios. The brightness temperatures predicted by both the LTE and non-LTE models are shown in Figure~\ref{tempsone}, emphasizing the large deviations beween the predictions of the two models, especially for lower $J$ transitions.  The measured data points have been scaled by the lensing magnification factors listed in Table~\ref{tab:cont} when available, and by a factor of 10 in all other cases. This figure shows that the constraints on the model parameters can be tightened by measurements of lower $J$ transitions, especially the CO(1$-$0) line. Even if most of the CO(1$-$0) emission comes from a colder gas component, using this value as an upper limit will help rule out some regions of the parameter space, as in our example for SDP.81. 

The properties of the LTE and non-LTE models could be compared by calculating the $total$ CO luminosity, summed over all transitions in the model, which is correlated to the star formation rate and efficiency. Since the brightness temperature of the CO(1$-$0) line tends to be lower in the LTE models, translating into a lower total gas mass, the star formation efficiency, quantified by the $L_{IR}$/$L_{CO}^{'}$ ratio, will be higher in this case. As the temperature of the gas increases, more of the rotational CO lines become optically thick and high-J transitions start to dominate the gas cooling. Since the dominant cooling CO line is temperature-dependent, the total CO luminosity will be in general a better proxy for the total cooling rate than the luminosity of a particular transition. \citet{Bayet2009} also find a strong correlation between the total $L_{CO}^{'}$ and $L_{IR}$, using a mixed sample of nearby and high redshift galaxies. If this relationship holds, we find that both LTE and non-LTE models are overpredicting the measured $L_{IR}$, with a larger discrepancy in the non-LTE case. However, in both cases the total $L_{CO}^{'}$ integrated over all lines is still consistent within the scatter with the \citet{Bayet2009} correlation, and therefore we cannot rule out either scenario based on this comparison.

Distinguishing between the different regions in parameter space will clarify the state of the ISM in these galaxies, and thus their star formation histories. Specifically, hot/low-density gas may signal the action of a feedback process on star formation, increasing the Jeans mass. Other studies of high redshift SMGs find a warm CO component with $n$[H$_{2}$] around 10$^{4}$~cm$^{-3}$ and temperatures between $\sim$40 and 60~K \citep{Riechers2010,Carilli2010,Danielson2010}, a region marginally allowed by our contours. However, a direct comparison with results obtained from SLEDs extending down to CO(2$-$1) becomes less warranted in view of increasing evidence \citep[e.g., ][]{Panuzzo2010,bradford09} that the mid- and high-J CO lines are originating in some cases from a hot gas component. This high temperature/low density solution has not been fully investigated, but recent studies show that other CO SLEDs can be consistent with it \citep{scott10,Panuzzo2010,Weiss2007,Ao2008,Bayet2009}. The CO SLED in M82 is fit by a low-mass ($\sim$10\% of the total) CO component with a kinetic temperature of almost 600~K \citep{Panuzzo2010}, while solutions with $T_{kin}$ of a few$\times$100~K are found by \citet{scott10,Bayet2009}, and can be allowed by the LVG models for IRAS F10212+4724 \citep{Ao2008} and APM 08279+5255 \citep{Weiss2007}. Similarly, the Herschel-SPIRE spectrum of Arp~220 shows that the mid-$J$ CO luminosity is dominated by a gas component with T$\sim$1350~K, which represents only $\sim$10\% of the total CO mass \citep{Rangwala2011}. Such temperatures suggest energy input from outflows or AGN activity. The presence of an AGN component in SDP.17b is supported by the relatively flat SLED from CO(6$-$5) to CO(8$-$7), similar to the Cloverleaf quasar or Mrk231 \citep{bradford09,vanderWerf2010}, and the emission line of water, also observed in galaxies with an AGN component, such as Mrk231 \citep{Gonzales2010}.

\section{CONCLUSIONS}\label{sconcl}

Far-IR / submillimeter-wave surveys are revealing submillimeter-bright galaxies from the first half of the history of the Universe by the tens of thousands, but their detailed study requires spectroscopic redshift measurements. We have studied a sample of the brightest sources and have demonstrated a new redshift-measurement technique with our broadband millimeter-wave grating spectrometer, Z-Spec. Z-Spec measures multiple rotational transitions of carbon monoxide, a major coolant of molecular gas in galaxies, and thus is not dependent on optical counterparts which are often absent or hard to identify, as is the case for these galaxies. We find redshifts ranging roughly between 1 and 3, reaching back to an era when the Universe was 15\% of its present age. Their fluxes are proven to be amplified by gravitational lensing \citep{Negrello10}, making them ideal targets for spectroscopic follow-ups. From the observed CO line luminosities and integrated $L_{IR}$, typical conversion factors reveal that these galaxies each house roughly 10$^{10}$ M$_{\odot}$ of molecular gas, and have SFRs between 10$^{2}$ and 10$^{3}$~M$_{\odot}$yr$^{-1}$, after correcting for lensing magnification. Regardless of the magnification details, we are clearly witnessing a rare episode of rapid star formation in these galaxies, since the timescale over which the observed luminosity can be generated by converting the inferred mass of gas into stars is only a few tens of millions of years (depending on the details of the star formation and the accretion of more gas), which is a small fraction of the Universe's age even at this early epoch. We estimate that the dust masses in our sample of lensed galaxies are around a few$\times$10$^{8}$~M$_{\odot}$, and the wavelengths corresponding to the peaks of their dust SEDs fall within a narrow range, between 73 and 92~$\mu$m in the rest frame. For this initial set of lensed submm galaxies both the dust properties derived from the IR SED, and the physical conditions of the molecular gas probed by the CO lines, are broadly comparable to those in known SMGs  \citep{greve05,Solomon2005,Casey09}, with excitation temperatures in the 30-120~K range, and $L_{CO}^{'}$/$L_{IR}$ between 1 and 3$\times$10$^{-3}$ K~km~s$^{-1}$~pc$^{2}$/L$_{\odot}$, as measured from the mid-$J$ CO lines. 

The partial SLEDs for the CO molecule constructed from the lines observed by Z-Spec cannot distinguish between different models of CO excitation. The simplest assumption is that of local thermodynamic equilibrium (LTE), under which we can derive the gas column density and excitation temperature. We find that the relative line strengths can be reproduced by relatively low excitation temperatures ($<$~100~K), and optical depths ($<$~1). In the non-LTE case, other parts of the parameter space are allowed, including higher optical depths, while measurements of the lower rotational transitions are essential in confirming such models.

By being able to characterize galaxies that can be inaccessible at other wavelengths, the combination of large-area submm surveys and spectroscopic follow-ups of the CO emission lines will lead to substantial progress in our understanding of high redshift galaxies and their evolution.
These results suggest the possibility of a rapid growth in our understanding of high redshift star formation in highly dust-obscured galaxies, independent of identifying optical or radio counterparts, but enabled by strong gravitational lensing magnification. 

\acknowledgements
We are indebted to the staff of the Caltech Submillimeter Observatory for their unflagging support. This work was supported by NSF grant AST-0807990 to J. Aguirre and by the CSO NSF Cooperative Agreement AST-0838261. Support was provided to J. Kamenetzky by an NSF Graduate Research Fellowship. Z-spec was constructed under NASA SARA grants NAGS-11911 and NAGS-12788 and an NSF Career grant (AST-0239270) and a Research Corporation Award (RI0928) to J. Glenn, in collaboration with the Jet Propulsion Laboratory, California Institute of Technology, under a contract with the National Aeronautics and Space Administration. We acknowledge Peter Ade and his group for their filters and Lionel Duband for the 3He / 4He refrigerator in Z-Spec, and are grateful for their help in the early integration of the instrument. R.L. wishes to thank Tom Loredo for useful discussions regarding the significance of the redshift determination, and P. Papadopoulos for help improving the gas mass discussion. R.L. also thanks the anonymous referee for the thorough read and the helpful comments for improving the paper. We appreciate the help of Robert Hanni and Jon Rodriguez with observing.

\begin{appendix}
\section{}

Our redshift determination is based on defining the probability of false positives (or the false detection rate - FDR) in the absence of signal, and choosing the combination of estimators that produces the lowest FDR, and therefore the largest significance. In order to justify these choices and definitions, as well as the $\sqrt{N}$ normalization factor for $E_2$, we will characterize and compare the distributions of the estimators, given that the signal $S_i$ in each channel is a normal random variable. This is the assumption of our simulations, which lead to the definition of the redshift significance. We verify that the distributions of the estimators are constant over the redshift range considered ($0.5-6$).

\subsection{Gaussianity}
All three distributions, denoted $f(E_1(z))$, $f(E_2(z))$, and $f(E_3(z))$, respectively, are gaussian. This can be easily verified for $f(E_1)$ and $f(E_3)$, since by definition they are constructed as linear combinations of normal random variables. $E_2$ on the other hand, is defined, up to the normalization constant, as a sample median, which is a central order statistics. Numerous results \citep[see ][and references therein]{Shorack1973, Ruymgaart1977, Mason1992} show that order statistics, as well as the linear combinations of order statistics, of i.i.d. (independent and identically distributed) and non-i.i.d. variables are asymptotically normal. These conditions apply for the sample sizes $n\sim10^6$ of our simulations, and therefore $f(E_2)$ will also be well described by a gaussian distribution.
 
 \subsection{Expected values}
The expected values of the estimators should be 0 at all redshifts for noise spectra, and should be largest at the correct redshift when lines are present in the spectrum. Taken independently or jointly, the values of these estimators determine the significance of the identified redshift.

Let $N(z)$ denote the number of CO lines falling in the Z-Spec bandpass at redshift $z$. In the {\it noise simulations}, for each channel $i$, $1\leq i\leq N(z)$, the signal values $S_i$ are drawn from a normal distribution, with mean 0, and standard deviation equal to the noise value, $\sigma_i$. Therefore, all $S_i/\sigma_i$ will be distributed as $\mathcal{N}(0,1)$. 

It's easy to see that all our estimators have an expected value of 0 in the absence of signal. In this case, $E_1$ and $E_3$ are just linear combinations of i.i.d. normal variables with mean 0. For simplicity, let's denote $S_{i}/\sigma_{i}=x_i$, where the $x_i$'s are i.i.d. $\mathcal{N}(0,1)$, and re-write the definition of $E_2$ as

\begin{equation}\label{a1eq}
E_{2}(z)=\sqrt{N(z)}\times\mathrm{median}(\mathcal{A}),
\end{equation}
where $\mathcal{A}$ denotes the set
 
\begin{equation}\label{a2eq}
\mathcal{A}=\{ f_{ij} | f_{ij}=0.5(x_i+x_j), 1\leq i,j\leq N(z), i < j\}.
\end{equation}
 
\noindent The set $\mathcal{A}$ has $M$ elements, with $M=N(N-1)/2$, and each element $f_ij=0.5(x_i+x_j)$ will be a $\mathcal{N}(0,1/2)$ random variable. Since the expected value of the sample median is equal to the median of the underlying distribution, $\mathcal{N}(0,1/2)$, which is also 0 (for a gaussian, the median is equal to the mean), the noise distribution of $E_2$ is also a gaussian with mean 0.
 
If lines are present, let us assume that all $S_i$ (i.e. all channels containing a line) have the same mean $S_0$, and therefore are distributed as $\mathcal{N}(S_0,\sigma_i^2)$. This is a simplifying assumption, which leads to a straightforward comparison of the estimators. In this case, we can also write the signal as $S_{i}=S_0+\delta S_i$, where $\delta S_i$ are $\mathcal{N}(0,\sigma_i^2)$. In this case however, the $S_i/\sigma_i$ ratios will no longer be identically distributed, each having a normal distribution with a different mean, $\mathcal{N}(S_0/\sigma_i,1)$. From the definitions given in Section~\ref{sec:alg}, we have for the expected values of the estimators:

\begin{equation}\label{a3eq}
\mathcal{E}(E_{1}(z))=\frac{NS_{0}}{\sqrt{\sum_{i} \sigma_{i}^{2}}}=\frac{\sqrt{N}S_0}{\sqrt{<\sigma^2>}},
\end{equation}
 
and

\begin{equation}\label{a4eq}
\mathcal{E}(E_{3}(z))=\frac{S_0}{\sqrt{N}}\sum_{i} {\frac{1}{\sigma_{i}}}=\sqrt{N}S_0<\frac{1}{\sigma}>,
\end{equation}
 
 where
 
\begin{equation}\label{a5eq}
\begin{split}
<\frac{1}{\sigma}>=&\frac{1}{N}\sum_{i}{\frac{1}{\sigma_{i}}}, \\
<\sigma^2>= &\frac{1}{N}\sum_{i}{\sigma_{i}^2}.
\end{split}
\end{equation}
 
\noindent Note that the expected values of the estimators are calculated over all simulations, while the average of the $\sigma_i$'s is taken over the set of $N(z)$ lines observed at redshift $z$. Applying the Cauchy-Schwarz inequality $(<ab>^2\leq<a^2><b^2>)$, we have that $<\sigma^2>\geq<\sigma>^2$ and $\sqrt{<\sigma^2>}\geq<\sigma>$, and also $1\leq<\sigma><1/\sigma>$ so that $1/<\sigma>\leq <1/\sigma>$. This translates into 

\begin{equation}\label{a6eq}
\mathcal{E}(E_{1}(z))=\frac{\sqrt{N}S_0}{\sqrt{<\sigma^2>}}\leq\frac{\sqrt{N}S_0}{<\sigma>}\leq\sqrt{N}S_0<\frac{1}{\sigma}>=\mathcal{E}(E_{3}(z)).
\end{equation}
 
\noindent This proves that $E_3$ has a larger expected value than $E_1$, and will lead to a higher significance result than $E_1$, when used independently. 

The expected value of $E_2$ depends not just on some average of the noise values in all the channels where the lines fall, like $E_1$ and $E_3$, but on the $distribution$ of these noise values. Depending on whether $M$ is odd or even, either 2, 3, or 4 channels will determine the value of $median(\mathcal{A})$, and, depending on the noise distribution over the Z-Spec channels, the average noise in this subset of channels could be either larger or smaller than the average noise of $all$ the channels containing lines. Due to this distribution of noise among the used channels, which is dependent on $z$, $\mathcal{E}(E_2)$ is not consistently larger or smaller than $\mathcal{E}(E_3)$ for any value of $z$, but we can set a limit for $|\mathcal{E}(E_3)-\mathcal{E}(E_2)|$, independent of the values of $S_i$ and $\sigma_i$, and therefore independent of $z$.

By Chebyshev's inequality, the distance between the mean and the median is always less than, or equal to the standard deviation, $|mean(\mathcal{A})-median(\mathcal{A})|\leq \sigma_{\mathcal{A}}$. On the other hand, $\sqrt{N(z)}mean(\mathcal{A})$ is equal to $E_3$:
 
\begin{equation}\label{a7eq}
\sqrt{N}\frac{1}{M}\sum_{ij}{0.5(x_i+x_j)}=\sqrt{N}\frac{2}{2N(N-1)}(N-1)\sum_{i}{x_i}=\frac{1}{\sqrt{N}}\sum_{i}\frac{S_{i}}{\sigma_{i}},
 \end{equation}

\noindent since each $x_i$ will appear $N-1$ times in the sum. The equality holds for any distribution of the signal-to-noise. As a consequence, the standard deviation of the sample $mean$ for $\sqrt{N(z)}\mathcal{A}$ will also be equal to the standard deviation of $E_3$. Intuitively, the normalization factor used for $E_2$, $\sqrt{N(z)}$, allows us to apply Chebyshev's inequality in this way.

For the standard deviation of $\mathcal{A}$ we have

\begin{equation}\label{a8eq}
\sigma^2_{\mathcal{A}}=\frac{1}{2(N+1)}\sum_{i}{(x_i-\bar{x})^2}=\frac{N-1}{2(N+1)}s^2(x_i)=\frac{N-1}{2(N+1)}S_0^2s^2(1/\sigma_i),
 \end{equation}

\noindent where $s(x_i)$ denotes the {\it sample standard deviation}, and $\bar{x}$ denotes the {\it sample mean}. Combining Equations~\ref{a7eq} and \ref{a8eq}, we obtain the constraint

\begin{equation}\label{a9eq}
(E_{3}(z)-E_{2}(z))^2 \leq \frac{N(N-1)}{2(N+1)}S_0^2s^2(1/\sigma_i),
 \end{equation}
 
\noindent which shows that, since the sample $\mathcal{A}$ has a lower standard deviation than the original set $\{S_i/\sigma_i\}$ ($(N-1)/2(N+1) < 1$), the $median$ of $\mathcal{A}$ will be closer than the $median$ of $\{S_i/\sigma_i\}$ to the average value of ${S_i/\sigma_i}$. Therefore, $E_2$ could be a better estimator than $E_3$ for the case in which only a few of the $N(z)$ channels are noisier than average (i.e. few outliers). 


\begin {figure}[h]
\begin{minipage}[b]{0.48\linewidth}
\centering
\includegraphics*[scale=0.40,angle=0]{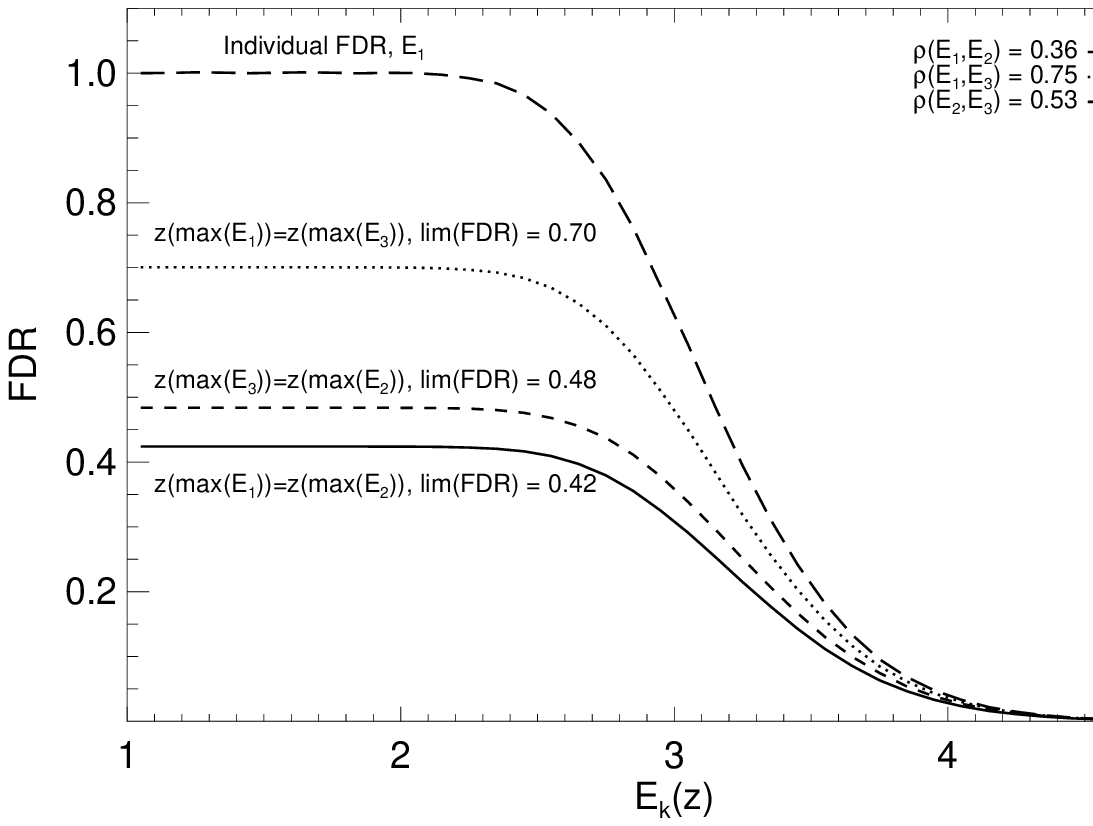}
\caption{Comparison of the FDR's as a function of the pair of estimators selected. The long-dash line shows the FDR for a single estimator, $E_1$. The notation $E_k$ corresponds to $E_1$ for the $(E_1,E_2)$ and $(E_1,E_3)$ pairs, respectively $E_2$ for the $(E_2,E_3)$ pair. On the upper right corner we list the values of the Pearson correlation coefficient for the same pairs of estimators. Note that the derived FDR decreases as the correlation between estimators decreases. The $(E_1,E_2)$ pair has the lowest correlation and also leads to the lowest FDR. \label{ecorr}}
\end{minipage}
\hspace{0.1in}
\begin{minipage}[b]{0.48\linewidth}
\centering
\includegraphics*[scale=0.50,angle=90]{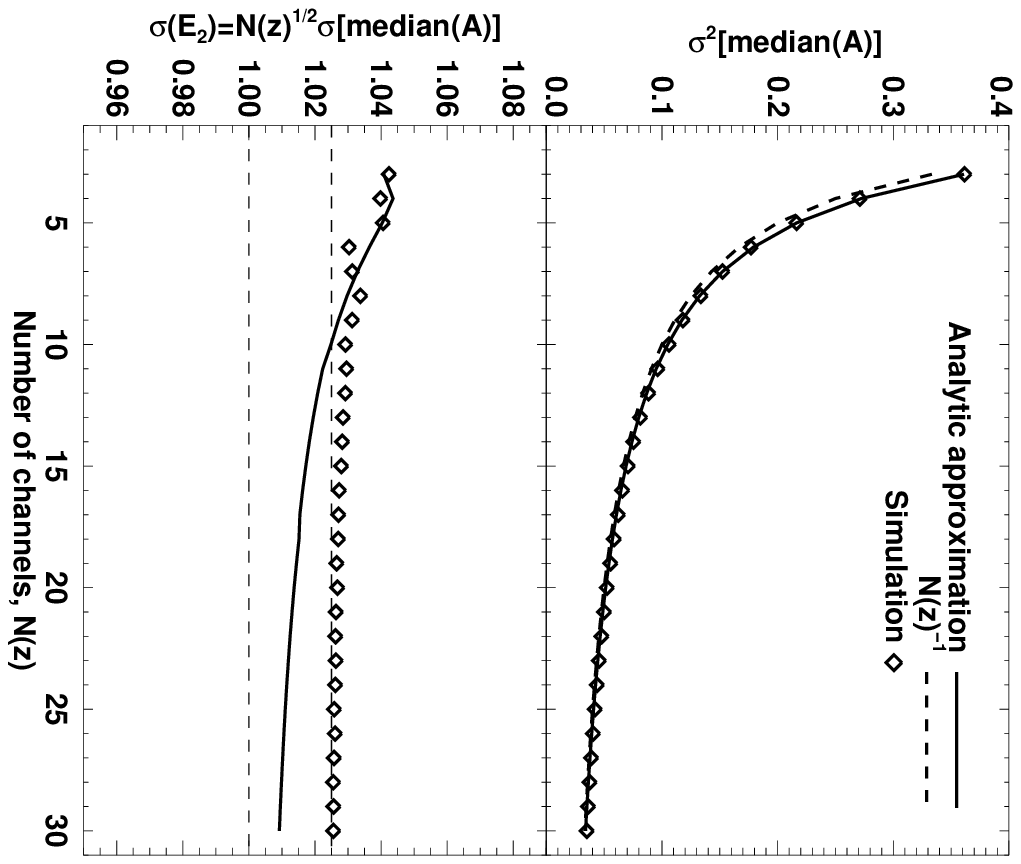}
\caption{Upper panel: variance of the median of set $\mathcal{A}$, calculated using the approximation in Equation~\ref{a18eq}, with the correction for small $N$ (solid line), and obtained by simulations (diamonds). For comparison, the dashed line shows the $1/N$ curve. Lower panel: the standard deviation of $E_2$, obtained by normalizing $median(\mathcal{A})$ by the $\sqrt{N}$ factor. The solid line and the diamonds represent the analytic approximation, and the simulations, respectively, while the dotted lines are plotted to guide the eye. Note that the simulated points are in fact more linear than the semi-analytic formula, supporting an uniform $\sqrt{N}$ normalization factor, and the deviations from unity are only a few percent. \label{esigma}}
\end{minipage}
\end{figure}

The choice of any single estimator would be motivated by these individual properties. However, in the attempt to reduce the false detection rate even further, we combine these estimators in pairs, by requiring that their maxima occur at the same redshift. Figure~\ref{ecorr} shows the combined FDR obtained for each of the three pairs of estimators, as a function of the estimator value ($E_1$ for the $(E_1, E_2)$ and $(E_1, E_3)$ pairs, and $E_2$ for the $(E_2, E_3)$ pair). The FDR of a single estimator is also plotted with a long-dashed line, showing that for the same values of the estimator maxima, the FDR is lower in the combined case than in an individual case. The values of the Pearson correlation coefficients between these estimators, listed in the upper right corner, show that a lower correlation is associated with a lower FDR, since in this case the maxima are less likely to occur at the same redshift. The $(E_1,E_2)$ pair is the one with the lowest correlation and FDR, and is the one used further in our algorithm.

\subsection{Variance}
 
The last step in characterizing the properties of our estimators is deriving their variance. We will show that $E_1$ and $E_3$ have a variance equal to 1, while the normalization factor for $E_2$ also brings is variance within a few precent of this value.

Regardless of the expected value of the signal per channel, $S_0$,  $Var(E_{3}(z))=Var(E_{1}(z))=1$, since 

\begin{equation}\label{a10eq}
\begin{split}
\mathcal{E}(E_{1}^2(z))= &1+\frac{NS_0^2}{<\sigma^2>},\\
\mathcal{E}(E_{3}^2(z))=& 1+NS_0^2<\frac{1}{\sigma}>^2,
\end{split}
\end{equation}

\noindent and $Var(X)=\mathcal{E}(X^2)-\mathcal{E}(X)^2$, where $\mathcal{E}(X)$ is given by Equations~\ref{a3eq} and \ref{a4eq} for the two estimators, respectively. For the noise distribution, the variance of $E_{3}(z)$ also follows immediately from the fact that the variance of the sample mean of $N$ i.i.d. $\mathcal{N}(\mu_0,\sigma_0^2)$ random variables is $Var(\bar{X})=\sigma^2_0/N$:

\begin{equation}\label{a11eq}
Var(E_{3})=\frac{1}{N}Var\left(\sum_{i}\frac{S_{i}}{\sigma_{i}}\right)=NVar\left(\frac{1}{N}\sum_{i}\frac{S_{i}}{\sigma_{i}}\right)=1.
\end{equation}

\noindent It is trivial to see that for $N=2$, the variance of $E_2$ is also 1, since in this case $E_2=E_3$. For larger values of $N$, the presence of correlations among the elements of $\mathcal{A}$ introduces important complications in deriving an analytic form. In fact, the covariance between any $(f_ij,f_ik)=(0.5(x_i+x_j),0.5(x_i+x_k))$ pair is 1/4, and each $f_ij=(x_i+x_j)/2$ is correlated with $2N-4$ other variables. Therefore, the covariance matrix of $\mathcal{A}$ will have 1/2 on the diagonal, $N(N-1)(N-2)$ elements equal to 1/4, and the rest will be 0's.

Up to the normalization factor, $E_2$ is defined as the median of the set $\mathcal{A}$, which is the central order statistic when $M=N(N-1)/2$ is odd, and a linear combination of order statistics when M is even (the average of the two values in the middle). Analytic expressions for the moments of order statistics have been derived for the case of i.i.d. normal random variables \citep{David2003,Tong1990}, and generalized in a simple form only for the case of non-i.i.d. $exchangeable$ normal random variables \citep{Owen1962,Tong1990}. Exchangeable random variables are equicorrelated: the correlation matrix has all off-diagonal elements equal. However, the variables contained in the set $\mathcal{A}$ are non-i.i.d., and are $not$ equicorrelated, except in the case when $N=3$. For noise simulations, the $f_ij$'s are drawn from the same distribution, but are correlated (ni.i.d), while if lines are present, the $f_ij$'s will also have different means, becoming also non-identically distributed (ni.ni.d.). No analytic expressions for the moments exist for this case, and only a few general relations between the order statistics of such non-i.i.d. variables have been established \citep{Balakri1992,Tong1990}. We can show, however, that the variance of the median of $\mathcal{A}$ can be approximated analytically, and justify the choice of the normalization factor for $E_2$.

For a sample of i.i.d. normal random variables, it is well known \citep{Cramer1946} that the sample median has an asymptotically normal distribution, with variance 

\begin{equation}\label{a12eq}
Var_{iid}(\tilde{X}_M)\rightarrow\frac{1}{4f(\tilde{X})^2M}=\frac{\pi\sigma_0^2}{2M},
\end{equation}

\noindent where $\tilde{X}_M$ denotes the sample median, $f(\tilde{X})$ is the value of the distribution function at the position of the median, and $M$ is the sample size. For a normal distribution $\mathcal{N}(\mu_0,\sigma_0^2)$, $f(\tilde{X})=1/\sqrt{2\pi\sigma_0^2}$. Except for this asymptotic case, there are no analytic forms for the moments of the sample median for the normal distribution, and the integrations have to be performed numerically. The value of $Var_{iid}(\tilde{X}_M)$ has been tabulated in the literature \citep{Teichroew1956,Tietjen1977}. We use the approximation

\begin{equation}\label{a13eq}
Var_{iid}(\tilde{X}_M)\approx\frac{4^{2m}m!^4}{(2m+1)!^22\pi f(\tilde{X})^2}=\frac{4^{2m}m!^4}{(2m+1)!^2},
\end{equation}

\noindent based on the coefficient multiplying the exponential part of the distribution function, with $m$ defined as $M=2m+1$. Empirically, this expression offers a better approximation for small $m$'s, than the one derived from the exponent of the exponential (given in Equation \ref{a12eq}).The last equality in Equation~\ref{a13eq} follows for an $\mathcal{N}(0,1)$ distribution. 

For a sample of non-i.i.d. $equicorrelated$ random variables, with the same mean $\mu_0$, same variance $\sigma_0^2$, and same covariance $C_0$, \citet{Owen1962} showed that the variance of the median can be witten as

\begin{equation}\label{a14eq}
Var_{niid}(\tilde{X}_M)\approx C_0+(\sigma_0^2-C_0)Var_{iid}(\tilde{X}_M)
\end{equation}

\noindent where $Var_{iid}(\tilde{X}_M)$ is the variance for the sample median of $M$ i.i.d. random variables distributed as $\mathcal{N}(0,1)$. This relation provides an exact solution for the variance of $E_2$ in the case $N=M=3$. Since the elements of $\mathcal{A}$ are $\mathcal{N}(0,1/2)$, and the covariance for any correlated pair is $C_{\mathcal{A}}=1/4$, from Equation~\ref{a14eq} follows that

\begin{equation}\label{a15eq}
Var_{\mathcal{A}}(\tilde{X}_{M=3})=\frac{1}{4}+(\frac{1}{2}-\frac{1}{4})var_{iid}(\tilde{X}_M)\approx 0.361 \approx 1.08/N ,
\end{equation}

\noindent where we used Equation~\ref{a13eq} as an approximation for small $M$.

While an exact solution to the problem of non-equicorrelated variables is in principle possible \citep{Rawlings1976,Hill1976}, it involves a large number of integrations that ultimately have to be performed numerically. A similar situation arises in the study of genetic inheritance, and has led several authors \citep{Meuwissen1991,Phocas1995} to develop approximate solutions, by assuming that all the variables are equicorrelated, with a correlation coefficient equal to their $average$ correlation, and further refining this approximation using polynomial fits to Monte Carlo simulations.

Following \citet{Phocas1995}, we can define an average covariance $C_{eff}$, and assume that the sample $\mathcal{A}$ will behave as an equicorrelated sample with the new "effective" correlation. By definition, the variance of the sample $mean$ is

\begin{equation}\label{a16eq}
Var(\bar{X})=\frac{1}{M^2}\sum_i{Var(X_i)}+\frac{2}{M^2}\sum_{i,j>i}{Cov(X_i,X_j)}.
\end{equation}

\noindent We define the average covariance, $C_{eff}$, as the value that summed over all pairs produces the same total covariance. Since the covariance sum has $M(M-1)/2$ terms, we have

\begin{equation}\label{a17eq}
Var(\bar{X})=\frac{M\sigma_0^2}{M^2}+\frac{2}{M^2}\left(\frac{M(M-1)}{2}C_{eff}\right)=\frac{\sigma_0^2}{M}+\frac{M-1}{M}C_{eff},
\end{equation}

\noindent In order to derive $C_{eff}$, let's remember that, since the mean of sample $\mathcal{A}$ is equal to $E_3/\sqrt{N}$ (by Equation~\ref{a7eq}), it will also have the same variance as $E_3/\sqrt{N}$, namely $1/N$. By equating Equation~\ref{a17eq} with $1/N$, and taking into account that $\sigma_0^2=1/2$, after some algebra we obtain $C_{eff}=1/(N+1)$. Substituting $C_{eff}$ for $C_0$ in Equation~\ref{a14eq}, the expression for the variance of the median for the sample $\mathcal{A}$ becomes

\begin{equation}\label{a18eq}
Var_{\mathcal{A}}(\tilde{X}_M)\approx\frac{1}{N+1}+\frac{N-1}{2(N+1)}Var_{iid}(\tilde{X}_M)=\frac{1}{N+1}+\frac{N-1}{2(N+1)}\frac{\pi}{(N+1)(N-2)},
\end{equation}

\noindent where the last equality holds for large $N$'s and can be approximated as $1/N +0.57/N^2$, and therefore has an overall $1/N$ behavior. For small $N$'s, we calculate the values of $Var_{\mathcal{A}}(\tilde{X}_M)$ numerically,  replacing $Var_{iid}(\tilde{X}_M)$ by Equation~\ref{a13eq}. In this case, $Var_{\mathcal{A}}(\tilde{X}_M)=1/N +\mathcal{O}(1/N^k)$, where $\mathcal{O}(1/N^k)$ has values less than 0.03. Thus, by semi-analytic arguments, $Var_{\mathcal{A}}(\tilde{X}_M)\approx1/N$.

Even without additional polynomial corrections, the expression in Equation~\ref{a18eq} reproduces the actual variance within a few percent. We have checked this result by numerical simulations, obtained by drawing $N(z)$ numbers from a $\mathcal{N}(0,1)$ distribution (corresponding to the $S_i/\sigma_i$ variables for the noise spectrum), constructing the set $\mathcal{A}$, and taking its median. The expected value and variance of $Var_{\mathcal{A}}(\tilde{X}_M)$ for each $N(z)$ have been calculated from $10^6$ such samples. The upper panel Figure~\ref{esigma} shows the analytic approximation with a solid line, and the simulated points as diamonds. The $1/N$ dependence is overplotted with a dashed line. For the analytic curve we have used the expression in Equation~\ref{a13eq} for $N<15$ and Equation~\ref{a18eq} for larger $N$'s. The bottom panel of the same figure shows a better comparison, obtained by multiplying the same curve and the points by $N(z)$ and taking the square root. In this case, the values plotted represent the standard deviation of $E_2$. From this figure it is apparent that, while our analytic approximation reproduces the behavior of $Var(E_2)$ within a few percent, based on the numerical simulations $Var(E_2)$ is in fact even more linear than the approximation suggests, which further justifies the choice of the $\sqrt{N}$ as the normalization constant for $E_2$. Even if the simulations do not asymptote exactly to 1, the difference is a constant multiplication factor, showing that there is no additional $N$ dependence.

\end{appendix}
\vspace{0.3in}
\bibliography{references}

\end{document}